\documentclass[final,journal,twoside,twocolumn,10pt]{IEEEtran}
\usepackage{acronym}
\usepackage{amsmath}
\usepackage{graphicx}
\usepackage{hyperref}
\usepackage{xcolor}

\newcommand{\bracket}[1]{\begin{bmatrix} #1 \end{bmatrix}}
\newcommand{\insertpng}[1]{\includegraphics[width=\textwidth]{#1}}
\newcommand{\insertpdf}[1]{\includegraphics[scale=0.65]{#1}}
\newcommand{\mycaption}[2]{\caption[#1]{#1 #2}}

\newcommand{\B}{\mathrm{B}}
\newcommand{\GB}{\mathrm{GB}}
\newcommand{\s}{\mathrm{s}}
\newcommand{\T}{\mathrm{T}}

\DeclareMathOperator{\cat}{cat}
\DeclareMathOperator{\diag}{diag}
\DeclareMathOperator{\imag}{Im}
\DeclareMathOperator{\real}{Re}
\DeclareMathOperator{\tr}{tr}

\acrodef{CC0}{Creative Commons public domain}
\acrodef{CP}{canonical polyadic}
\acrodef{CTF}{Cyclops Tensor Framework}
\acrodef{DFT}{discrete Fourier transform}
\acrodef{EMV}{extended matrix-vector}
\acrodef{FFT}{fast Fourier transform}
\acrodef{HMF}{Hessian multiply function}
\acrodef{LFI}{light field imaging}
\acrodef{LHS}{left-hand side}
\acrodef{MDA}{multidimensional array}
\acrodef{MIA}{Multivariate Image Analysis}
\acrodef{MTT}{MATLAB Tensor Toolbox}
\acrodef{MV}{matrix-vector}
\acrodef{NT}{numeric tensor}
\acrodef{NWChem}{NorthWest Chemistry}
\acrodef{PC}{personal computer}
\acrodef{PLS}{Partial Least Squares}
\acrodef{RAM}{random access memory}
\acrodef{RHS}{right-hand side}
\acrodef{RT}{Ricci-notation tensor}
\acrodef{SDF}{structured data fusion}
\acrodef{SSE}{sum squared error}
\acrodef{SVD}{singular value decomposition}
\acrodef{TAMM}{Tensor Algebra for Many-Body Methods}
\acrodef{TCE}{Tensor Contraction Engine}
\acrodef{TensorLy}{Tensor Learning in Python}
\acrodef{TT}{tensor trains}

\begin{document}

%%%%%%%%%%%%%%%%%%%%%%%%%%%%%%%%%%
% Title and Authors
%%%%%%%%%%%%%%%%%%%%%%%%%%%%%%%%%%

\title{Ricci-Notation Tensor Framework for\\Model-based Approaches to Imaging}

\author{Dileepan~Joseph (dil.joseph@ualberta.ca)\\
\vspace{1ex}
Department of Electrical and Computer Engineering\\
University of Alberta, Edmonton, AB, Canada\\
\vspace{1ex}
June 4, 2024}

\date{June 4, 2024}
\maketitle

%%%%%%%%%%%%%%%%%%%%%%%%%%%%%%%%%%
% Abstract and Keywords
%%%%%%%%%%%%%%%%%%%%%%%%%%%%%%%%%%

\begin{abstract}
Model-based approaches to imaging, like specialized image enhancements in astronomy, facilitate explanations of relationships between observed inputs and computed outputs. These models may be expressed with extended matrix-vector (EMV) algebra, especially when they involve only scalars, vectors, and matrices, and with n-mode or index notations, when they involve multidimensional arrays, also called numeric tensors or, simply, tensors. While this paper features an example, inspired by exoplanet imaging, that employs tensors to reveal (inverse) 2D fast Fourier transforms in an image enhancement model, the work is actually about the tensor algebra and software, or tensor frameworks, available for model-based imaging. The paper proposes a Ricci-notation tensor (RT) framework, comprising a dual-variant index notation, with Einstein summation convention, and codesigned object-oriented software, called the RTToolbox for MATLAB. Extensions to Ricci notation offer novel representations for entrywise, pagewise, and broadcasting operations popular in EMV frameworks for imaging. Complementing the EMV algebra computable with MATLAB, the RTToolbox demonstrates programmatic and computational efficiency via careful design of numeric tensor and dual-variant index classes. Compared to its closest competitor, also a numeric tensor framework that uses index notation, the RT framework enables superior ways to model imaging problems and, thereby, to develop solutions.
\end{abstract}

%%%%%%%%%%%%%%%%%%%%%%%%%%%%%%%%%%
% Introduction
%%%%%%%%%%%%%%%%%%%%%%%%%%%%%%%%%%

\section{Introduction}
\label{sec:introduction}

``Tensors'' easily associate with learning-based approaches to imaging. After all, TensorFlow from Google Brain \cite{Abadi2016} and Tensor Comprehensions from Facebook AI \cite{Vasilache2018} are machine learning systems connected with a well known deep learning revolution in object detection and classification. Nonetheless, imaging researchers like Mai \emph{et al.} \cite{Mai2022} highlight the importance of model-based tensor approaches, emphasizing explanatory value. Mai \emph{et al.} also argue that a model-based approach helps to reduce the training data requirements for a learning-based approach, yielding what is in effect a hybrid approach.

Imaging researchers are finding advantages of model-based tensor approaches in a variety of applications. These include: multi-exposure, multi-focus, and hyper/multi-spectral image fusion to increase dynamic range, depth-of-field, and spatial resolution \cite{Mai2022, Wang2023, Prevost2022}; inpainting, or tensor completion, of missing or corrupt pixels in images and videos, even rows or columns of the same \cite{Liu2021, Bengua2017}; deblurring, or image restoration, posed as a constrained inverse problem often with added noise \cite{Newman2020, Lefkimmiatis2015, Rezghi2017}; compressive sensing, a mixture of computational imaging and image compression, for \ac{LFI}, synthetic aperature radar, and hyperspectral imaging modalities \cite{Marquez2020, Qiu2020, Li2020}; other constrained image enhancements, like ones that enforce 3D spherical invariances \cite{Skibbe2017}; and image quality assessments, like ones tailored for \ac{LFI} \cite{Zhou2020}.

Model-based imaging research involving tensors \cite{Mai2022}--\cite{Zhou2020} exhibits patterns. Papers typically have a section or subsection, on fundamentals, that addresses unary, binary, and $N$-ary operations expressible with tensors, like multi-way or multi-index contractions, Kronecker and Khatri-Rao products, and Tucker, \ac{CP}, and \ac{TT} constructs. More fundamental than operations, authors discuss an $n$-mode notation, especially for decomposition constructs, an index notation, also called Einstein notation, or a hybrid notation thereof. Sometimes, \ac{EMV} operators are employed to represent Hadamard, or entrywise, and Kronecker, or outer, products. Notation can resemble MATLAB or Python+NumPy syntax for \acp{MDA}. Instead of \ac{MV} or \ac{EMV} algebra, authors prefer tensor algebra to represent and manipulate high-order image relationships, including generalizations of the \ac{SVD} for rank reduction and feature extraction, akin to principal components analysis, and optimizations where images are the unknown variables in a functional that suffers upon vectorization of variables.

When literature on learning-based \cite{Abadi2016}--\cite{Vasilache2018} and model-based \cite{Mai2022}--\cite{Zhou2020} approaches to imaging use the word, ``tensor'' mainly or solely means a numeric data structure, usually an \ac{MDA}, together with numeric operations, like those identified with overloadable operators in object-oriented programming languages. This paper uses the word likewise, avoiding requirements for deeper tensor character addressed, for example by Synge and Schild \cite{Synge1978}, in books on geometric tensor calculus.

Whereas tensor notations and operations are in some senses implementation agnostic, a brief survey of tensor software is in order. Because its components, like the \ac{CTF} \cite{Solomonik2014, Singh2022}, may apply to tensor contractions, completions, and decompositions for imaging, the \ac{NWChem} software ecosystem is noteworthy \cite{Valiev2010, Kowalski2021}. With the \ac{TCE} and the \ac{TAMM} \cite{Baumgartner2005, Mutlu2023}, this ecosystem targets massively parallel computing and large-scale molecular simulations. Intended more for imaging and for desktop parallel computing, \ac{TensorLy} offers tensor decompositions to improve efficiency, robustness, and explainability of deep learning networks \cite{Kossaifi2019, Panagakis2021}. For desktop computing with MATLAB, the \ac{PLS} and \ac{MIA} Toolboxes, cited in biomedical imaging research \cite{Cumpson2015, Bedia2020}, offer tensor decompositions and imaging customizations.

As Harrison and Joseph argue \cite{Harrison2016}, the codesign of tensor algebra \emph{and} software, what they call a tensor \emph{framework}, has benefits. Codesign facilitates programmatic and computational efficiencies for solving a variety of model-based problems. In support, they cite Bader and Kolda's \cite{Bader2006, Kolda2009} development of $n$-mode+ notation and the \ac{MTT}, also called the Tensor Toolbox for MATLAB. While Bader and Kolda contributed notation for decomposition constructs along with original software, others leveraged existing notations while contributing decomposition constructs and MATLAB toolboxes. These include \ac{SDF} with Tensorlab, led by Sorber and De~Lathauwer \cite{Sorber2015, Vervliet2014}, and the TT-Toolbox with the \ac{TT} construct, by Osledets \cite{Oseledets2011}.

This paper proposes the \ac{RT} framework for model-based approaches to imaging. A successor to Harrison and Joseph's \ac{NT} framework \cite{Harrison2016}, itself a complement to an \ac{EMV} framework, the \ac{RT} framework inherits their advantages. The proposed framework comprises \ac{RT} algebra and \ac{RT} software, defined as the RTToolbox \cite{Joseph2023}, developed with this paper, plus MATLAB. The \ac{RT} software parses the \ac{RT} algebra, dispatching calculations to the MATLAB kernel, with no dependencies on other toolboxes.

Both the \ac{NT} and proposed \ac{RT} algebras are index notations that exploit and extend the Einstein summation convention. However, using a simpler dual-variant index notation, the \ac{RT} algebra proves equally expressive as the \ac{NT} algebra, shown here to be a multi-variant index notation, and aligns well with the Ricci notation of tensor calculus. Moreover, the \ac{RT} algebra supports additional outer operations. Inspired by the broadcasting of \ac{EMV} algebras, they remain grounded in a dual-variant index notation with summation convention.

The \ac{NT} software had options \cite{Harrison2016}. With LibNT, C/C++ end users expressed the \ac{NT} algebra with programmatic and, especially, computational efficiency. With NTToolbox, comprising M-file and MEX wrappers for LibNT, MATLAB end users expressed the \ac{NT} algebra with some programmatic and some computational penalties. The \ac{RT} software has only the RTToolbox option \cite{Joseph2023}, although this paper elaborates on its design in a C/C++-compatible way. Comprising just M-files, no compilation or linking is required. As per MATLAB Central recommendations \cite{MathWorks2022}, avoidance of MEX files and binaries means that sharing is easy via File Exchange.

For its computational efficiency, the \ac{RT} software leverages compiled unary and binary pagewise functions introduced, in 2020 and 2022, to the MATLAB kernel. Of note, the kernel \emph{pagewise} multiplication \cite{MathWorks2020} resembles the LibNT \emph{lattice} multiplication \cite{Harrison2016} introduced, in 2016, with the \ac{NT} software. This paper elaborates on the lattice concept to facilitate \ac{RT} left- and right-division using kernel functions. It also introduces pagewise concatenation, an interpreted function the \ac{RT} software exploits to support $N$-ary tensor concatenation. Programmatic efficiency of the \ac{RT} software is illustrated via software expressions cross-referenced to algebraic ones.

An image enhancement example, developed here, illustrates the \ac{RT} algebra. Inspired by Sirbu \emph{et al.}'s model-based research \cite{Sirbu2013, Sirbu2016, Sirbu2017} for coronagraphy \cite{Schulz2023}, the example adopts a simple coronagraph model with two planes related by 2D \acp{DFT} and a phase aberration. Starting with a \ac{CC0} image \cite{Wikipedia2022}, called \texttt{Airy\_disk\_D65}, the impact of the model on a synthetic ground-truth image having an occulted star with nearby exoplanets is computed. The exoplanets become visible after correction of the aberration, treated as unknown.

The rest of this paper is organized as follows. Section~\ref{sec:tensor algebra} introduces the \ac{RT} algebra, contrasting it with the \ac{NT} algebra. Section~\ref{sec:imaging example} illustrates the \ac{RT} algebra's usefulness for a model-based approach involving 2D \acp{DFT}. Section~\ref{sec:tensor software} summarizes the \ac{RT} software, contrasting it with the \ac{NT} software. Unlike Section~\ref{sec:tensor algebra}, Sections~\ref{sec:imaging example} and~\ref{sec:tensor software} include MATLAB code. Finally, Section~\ref{sec:conclusions} highlights contributions of this paper, this time with reference to selected details from Sections~\ref{sec:tensor algebra} to~\ref{sec:tensor software}.

%%%%%%%%%%%%%%%%%%%%%%%%%%%%%%%%%%
% Tensor Algebra
%%%%%%%%%%%%%%%%%%%%%%%%%%%%%%%%%%

\section{Tensor Algebra}
\label{sec:tensor algebra}

In the proposed \ac{RT} algebra, tensors are scalars, vectors, and matrices of any degree, where degree is the number of \emph{true}-variant and/or \emph{false}-variant indices. First, benefits of exploiting and extending Ricci notation for this purpose are introduced in the context of scalars, even though these benefits also apply to nonscalars. Second, additional extensions of the \ac{RT} algebra are summarized in the context of vectors and matrices.

\subsection{Dual-Variant Index Notation}
\label{sec:dual-variant index notation}

The proposed \ac{RT} algebra is introduced with reference to the \ac{NT} algebra. Using an underline operator in what its authors called a single-type index notation, the \ac{NT} algebra exploited and extended Einstein summation to support arbitrary mixtures of inner, entrywise, and outer products of $N$-degree scalars, not called $N$-dimensional arrays with reason. Use of the word degree (or order), instead of dimensional, for the number of indices is consistent with a geometric tensor calculus.

In the \ac{NT} algebra, the underline operator prevents Einstein summation, yielding entrywise products. One can demonstrate unnecessary complexity of the notation with simple examples. Consider the ternary inner product, $x$, and the ternary entrywise product, $y$, of three degree-one scalars, $a$, $b$, and $c$, with indices suppressed in inline math for readability:
\begin{align}
x &= a_i b_i c_i \text{,} \\
y_i &= a_{\underline i} b_{\underline i} c_{\underline i} \text{.}
\end{align}
For a variety of derivations, one needs to pair operands of $N$-ary products arbitrarily. This is done with commutation and association identities, as follows for these examples:
\begin{align}
x &= b_i (a_{\underline i} c_{\underline i}) \text{,} \\
y_i &= b_{\underline i} (a_{\underline {\underline i}} c_{\underline {\underline i}}) \text{.}
\end{align}
The underline operator stacks arbitrarily, meaning any index has a countable infinite number of variants. This freedom has implications for the \ac{NT} software. It also means the \ac{NT} algebra is actually a multi-type or multi-variant index notation.

Setting aside questions of contructivism, consider a simpler formalism, namely the \ac{RT} algebra, that enjoys the accepted advantages of the \ac{NT} algebra for numeric tensor purposes. With the \ac{RT} algebra, inspired by the Ricci notation of geometric tensor calculus, any index always has two variants, \emph{true} and \emph{false}, that are identified usually with typeset position, subscript and superscript, respectively. The \emph{co}variant and \emph{contra}variant indices of the Ricci notation motivated these names.

With the Ricci notation, an Einstein summation applies to each index that repeats twice where one is covariant and the other is contravariant. Repeating indices of the same variant is disallowed for geometric reasons. This convention specifies inner products or, for one operand, contraction (summation) along each repeated index. With the \ac{RT} algebra, indices of either variant need not be unique. Repeated indices of the same variant specify entrywise products or, for one operand, an attraction (selection) along each repeated index. Unique indices across multiple operands specify outer products.

To support $N$-ary expressions having any combination of inner, entrywise, and outer products, as opposed to a hierarchy of binary products indicated with parentheses or a left-to-right precedence rule, what matters for the \ac{RT} algebra is whether all repeats of an index do not or do involve the same variant. Returning to the ternary inner and entrywise examples, they are expressed and paired as follows, respectively:
\begin{align}
\label{eq:inner1}
x &= a_i b_i c^i \text{,} \\
&= b_i (a^i c^i) \text{,} \\
\label{eq:ewise}
y_i &= a_i b_i c_i \text{,} \\
&= b_i (a_i c_i) \text{.}
\end{align}
As with the \ac{NT} algebra, the \ac{RT} algebra enjoys full associativity and commutativity for $N$-ary products with scalar operands of arbitrary degree. With either algebra, the variant of one or more indices may change when altering associations.

Whereas inner products eliminate indices, tending to decrease the degree, outer products aggregate indices, tending to increase the degree. As they preserve repeated indices of the same variant, entrywise products tend to maintain the degree. Outer products are easily expressed as follows:
\begin{align}
\label{eq:outer1}
z_{ijk} &= a_i b_j c_k \text{.}
\end{align}
The examples of ternary inner, entrywise, and outer products are convenient for illustrating the known Ricci operation of contraction and an extended Ricci operation called attraction, both of which are unary operations that reduce degree:
\begin{align}
x &= z_{ii\bar{i}} \text{,} \\
y_i &= z_{iii} \text{.}
\end{align}
An alternative way, namely an overscript bar, to specify a false-variant index is shown. Unlike with the underline of the \ac{NT} algebra, each index of the \ac{RT} algebra still has two possible variants. Odd or even stacked overscript bars collapse into one or zero overscript bars. To specify a contraction, at least one repeated index must have a complementary variant.

With a superscript/subscript notation, often preferred in literature on geometric tensor calculus, ambiguity is possible in expressions where not just an index but a tensor repeats. Consider the following apparently nonsensical statement:
\begin{align}
y_i^j &\ne y_i^j \text{.}
\end{align}
With the overscript bar notation, one could instead write a sensible inequality to represent the asymmetry of a tensor:
\begin{align}
\label{eq:relate1}
y_{i\bar{j}} &\ne y_{\bar{j}i} \text{.}
\end{align}
Contrast this with the dot-spacer approach of Ricci notation, which allows superscripts for contravariant indices while clarifying index positions to resolve ambiguities as required:
\begin{align}
y_i^{.j} &\ne y_{.i}^j \text{.}
\end{align}

The \ac{RT} algebra allows the superscript/subscript notation, with dot spacers as required, and the proposed overscript bar notation, where all indices are subscripts and dot spacers are never required. Another case of relevance is when permutations matter. They generalize transpositions, even to scalars. Consider a permuted copy, $z$, of a degree-two scalar, $y$:
\begin{align}
\label{eq:permute}
z_i^{.j} &= y_{.i}^j \text{.}
\end{align}
Removing the dot spacers here results in a sensical statement that, however, does not specify a numerical permutation:
\begin{align}
z_i^j &= y_i^j \text{.}
\end{align}
The number of possible permutations grows with the factorial of degree. Context determines if they matter to a model.

Mixtures of inner, entrywise, and outer products, and ways in which to rearrange them, are significant to \ac{CP} derivations like those used to illustrate the \ac{NT} algebra. Consider the following key step of a degree-three \ac{CP} derivation, which uses the \ac{RT} overscript bar notation to clarify index positions and inner parentheses to group for asymptotic efficiency:
\begin{align}
u_{i\bar{\ell}} &= ((v_{j\ell} v_{\bar{j}\ell'}) (w_{k\ell} w_{\bar{k}\ell'})) \backslash (a_{ijk} v_{\bar{j}\ell'} w_{\bar{k}\ell'}) \text{.}
\end{align}
As with \ac{EMV} algebras, a left (or right) division implies the solution to a linear system. In this case, it is as follows:
\begin{align}
(v_{j\ell} v_{\bar{j}\ell'} w_{k\ell} w_{\bar{k}\ell'}) u_{i\bar{\ell}} &= a_{ijk} v_{\bar{j}\ell'} w_{\bar{k}\ell'} \text{.}
\end{align}
Iterating the key step and others like it, a degree-three scalar, $a$, decomposes into three degree-two factors, $u$, $v$, and $w$:
\begin{align}
a_{ijk} = \sigma_{\ell} u_{i\bar{\ell}} v_{j\bar{\ell}} w_{k\bar{\ell}} \text{,}
\end{align}
where the additional degree-one factor, $\sigma$, represents singular values, and where each degree-two factor obeys identities.

Previously, using the multi-variant index notation of the \ac{NT} algebra, a novel derivation of the \ac{CP} key step was provided. Because the same would be possible with the dual-variant index notation of the \ac{RT} algebra, consider instead a difference between the \ac{NT} and \ac{RT} algebras concerning division, in the context of a simplified version of the \ac{CP} key step:
\begin{align}
\label{eq:divide}
u_{i\bar{\ell}} &= A_{\ell\ell'} \backslash b_{i\ell'} \text{,} \\
\label{eq:mixed1}
A_{\ell\ell'} u_{i\bar{\ell}} &= b_{i\ell'} \text{.}
\end{align}
Provided that each index, in the denominator operand of a division, changes to its complement, the \ac{RT} algebra is equally consistent as the \ac{NT} algebra for implied linear systems.

With MATLAB's implied \ac{EMV} algebra, left/right division is equivalent to inversion or pseudo-inversion of the denominator operand followed by left/right multiplication. With the \ac{RT} algebra, as with the \ac{NT} algebra, there are multiple ways in which each index of a high-degree denominator operand may participate in a linear system, whether fully ranked or not. Absent the linear system context, fully specified by a left/right division, an inversion or pseudo-inversion of a denominator operand is ambiguous. Therefore, the \ac{RT} algebra favours binary division expressions over a complicated scheme presented with the \ac{NT} algebra to uniquely specify unary inverses.

In general, (partial) differentiation is explainable as the limiting solution of a linear system involving finite differences. The linear system may have an arbitrary mix of inner, entrywise, and outer products. Thus, (partial) differentiation rules for the \ac{RT} algebra inherit observations made in the context of linear systems. To ensure consistency in formulations involving differentials, each index in the denominator operand of a differential changes to its complement when evaluated.

\subsection{Nonscalars and Broadcasting}
\label{sec:nonscalars and broadcasting}

Like the \ac{NT} algebra, the \ac{RT} algebra supports nonscalars having indices, i.e., vectors and matrices of nonzero degree. In this paper, scalars are typeset with a plain italics font and nonscalars with a bold Roman font. When the degree is nonzero, indices are indicated in display math. Indices are suppressed in inline math where possible due to context.

In general, the product of two nonscalars is noncommutative, which limits associativity for $N$-ary products. Consider a degree-two product, $\mathbf{C}$, of degree-two matrices, $\mathbf{A}$ and $\mathbf{B}$, where columns of the first operand undergo an inner product with rows of the second operand, and where indices of the operands express entrywise products as per variants:
\begin{align}
\label{eq:mixed2}
\mathbf{C}_{i\bar{j}} &= \mathbf{A}_{i\bar{j}} \mathbf{B}_{i\bar{j}} \text{.}
\end{align}
As rows and columns, i.e., \ac{MV} indices, may convert to or from tensor indices, there is always an equivalent expression, with scalar operands of higher degree, that is fully commutative and associative for arbitrary product expressions, e.g.:
\begin{align}
c_{sti\bar{j}} &= a_{s\bar{r}i\bar{j}} b_{rti\bar{j}} \text{,} \\
&= b_{rti\bar{j}} a_{s\bar{r}i\bar{j}} \text{.}
\end{align}
Here, first and second indices of scalar operands correspond to rows and columns, respectively, of nonscalar operands. For an $N$-ary product of degree-two scalars, the operand sequence may be rearranged arbitrarily. This is not possible, in general, for an equivalent $N$-ary product of degree-zero matrices.

The binary product is reused to illustrate how a contraction or attraction is replaceable with an inner or entrywise operation across two operands with attention to index variants:
\begin{align}
\label{eq:contract}
t &= \tr \mathbf{C}_{i\bar{i}} \text{,} \\
&= \tr \mathbf{A}_{ii} \mathbf{B}_{\bar{i}\bar{i}} \text{,} \\
\label{eq:attract}
\mathbf{d}_i &= \diag \mathbf{C}_{ii} \text{,} \\
&= \diag \mathbf{A}_{ii} \mathbf{B}_{ii} \text{.}
\end{align}
Because of their resemblance to contraction and attraction, these examples include the trace operator, $\tr$, of \ac{MV} algebra, and the main-diagonal operator, $\diag$, of \ac{EMV} algebra, which concern only the rows and columns of an operand.

Inner products help justify why each enclosed index changes to its complement in a (conjugate) transposition. Consider these equivalent expressions for the Euclidean norm squared, $\|\mathbf{x}\|^2$, of an any-degree column vector, $\mathbf{x}$, having only false-variant indices, at first, and possibly-complex entries:
\begin{align}
\label{eq:inner2}
\|\mathbf{x}^\mathbf{k}\|^2 &= (\mathbf{x}^\mathbf{k})^* \mathbf{x}^\mathbf{k} \text{,} \\
&= (\bar{\mathbf{x}}^{k_1 k_2 \dots k_D})^\T \mathbf{x}^{k_1 k_2 \dots k_D} \text{,} \\
&= \bar{\mathbf{x}}_{k_1 k_2 \dots k_D}^\T \mathbf{x}^{k_1 k_2 \dots k_D} \text{,} \\
&= \mathbf{x}_\mathbf{k}^* \mathbf{x}^\mathbf{k} \text{.}
\end{align}
Here, indices of one variant in sequence are represented as an index vector. An overscript bar on a tensor, as opposed to an index, indicates complex conjugation of tensor entries.

Unlike the \ac{NT} algebra, the \ac{RT} algebra supports outer additions, subtractions, concatenations, and relations. The unifying idea is that any tensor of one degree may be expanded to an equivalent tensor of higher degree via an outer product with a unit tensor of the difference degree. Consider the following equations and relation, which consistently employ a degree-two matrix, $\mathbf{A}$, four degree-one column-vectors, $\mathbf{1}$, $\mathbf{b}$, $\mathbf{c}$, and $\mathbf{1}$, and an entrywise absolute-value operation, $|\cdot|$:
\begin{align}
\mathbf{A}_i^j &= \mathbf{1}_i (\mathbf{b}_j)^\T + |\mathbf{c}_i| (\mathbf{1}_j)^\T \text{,} \\
&= \bracket{\mathbf{1}_i & |\mathbf{c}_i|} \bracket{(\mathbf{b}_j)^\T \\ (\mathbf{1}_j)^\T} \text{,} \\
\mathbf{A}_i^j &\ge \mathbf{1}_i (\mathbf{b}_j)^\T \text{.}
\end{align}
Whereas the \ac{NT} algebra would require the with-unit outer products explicitly, as would Ricci notation, the \ac{RT} algebra allows them implicitly, an extension to Ricci notation:
\begin{align}
\mathbf{A}_i^j &= (\mathbf{b}_j)^\T + |\mathbf{c}_i| \text{,} \\
\label{eq:concat}
&= \bracket{\mathbf{1} & |\mathbf{c}_i|} \bracket{(\mathbf{b}_j)^\T \\ \mathbf{1}^\T} \text{,} \\
\label{eq:relate2}
\mathbf{A}_i^j &\ge (\mathbf{b}_j)^\T \text{.}
\end{align}
The missing outer products are implied by what may be called natural algebra, setting aside index variant details.

For an easier-to-read example, this time with scalars, of the naturalness of the proposed \ac{RT} algebra, consider a logarithmic model for the digital response or output, $y$, of a wide-dynamic-range image sensor to a uniform stimulus or input, $x$:
\begin{align}
\label{eq:outer2}
y_{ij} &= a_j + b_j \ln (c_j + x_i) + \epsilon_{ij} \text{.}
\end{align}
Offset, gain, and bias parameters, $a$, $b$, and $c$, depend on the pixel index, $j$, but not on the illumination index, $i$. In addition to such structural noise, the model accounts for temporal and quantization noise, $\epsilon$. A parseable expression of \ac{RT} algebra, the model implies three with-unit outer products, an entrywise function, and a mixed (entrywise-and-outer) product.

Outer \ac{RT} operations resemble broadcasting, popularized by the implied \ac{EMV} algebra of Python+NumPy, but are modelled with an index notation. Variants aside, one advantage of index-based outer operations is that they express broadcasting in a way that would look familiar to high-school students. This attests to the special value of the \ac{RT} algebra for model-based approaches, which favour human intelligence, in relation to learning-based approaches, which favour machine intelligence. By design, the \ac{RT} algebra includes and extends, with outer \ac{RT} operations, the implied \ac{EMV} algebra of MATLAB.

With MATLAB's implied \ac{EMV} algebra, an inner product of two matrices is representable as a product of vectors, with one conjugate transposed, after a vectorization is applied to each matrix. There is no MATLAB operator for an inner product of matrices, although there is one for an entrywise, or Hadamard, product of the same. In the \ac{RT} algebra, these operators may be denoted with a bullet and a circle, respectively.

Although \ac{EMV} algebra includes entrywise concatenation along non-row/column dimensions, computable in MATLAB via the kernel \texttt{cat} function, the \ac{RT} algebra offers a way to model it with tensor indices, a model generalizable to outer concatenation. Consider an outer concatenation of degree-two matrices, $\mathbf{A}$ and $\mathbf{B}$, along a common tensor index:
\begin{align}
\label{eq:indcat}
\mathbf{C}_{ijk} &= \cat_j \{\mathbf{A}_{ij}, \mathbf{B}_{jk}\} \text{,} \\
&= \delta_j^\ell (\alpha_\ell^j \mathbf{A}_{ij} + \beta_\ell^j \mathbf{B}_{jk}) \text{.}
\end{align}
The degree-three result, $\mathbf{C}$, is modelled with the help of degree-two scalars, $\alpha$ and $\beta$, each similar to a degree-two Dirac symbol, $\delta$, commonly used for index substitution:
\begin{align}
\alpha_\ell^j &= \begin{cases} 1, & \ell = j \text{,} \\ 0, & \text{otherwise} \text{,} \end{cases} \\
\beta_\ell^j &= \begin{cases} 1, & \ell = J_1 + j \text{,} \\ 0, & \text{otherwise} \text{.} \end{cases} 
\end{align}
The constant, $J_1$, is the dimension size of the tensor index, $j$, in the first operand, $\mathbf{A}$. With additional constants, $J_1 + J_2$, etc., the model is generalizable to $N$-ary concatenation.

%%%%%%%%%%%%%%%%%%%%%%%%%%%%%%%%%%
% Imaging Example
%%%%%%%%%%%%%%%%%%%%%%%%%%%%%%%%%%

\section{Imaging Example}
\label{sec:imaging example}

After proposing a problem, inspired by an exoplanet imaging instrument called a coronagraph, this section applies the \ac{RT} algebra and MATLAB to solve it. Results, presented first, summarize without equations or code a scalar optimization, called the phase aberration correction. Details, like a gradient matrix, are then modeled and simplified using (inverse) \ac{DFT} operators and the \ac{RT} algebra. Code, presented last, leverages (inverse) \acp{FFT} for efficiency.

\subsection{Simulation Results}
\label{sec:simulation results}

\begin{figure*}[t]
\centering
\insertpng{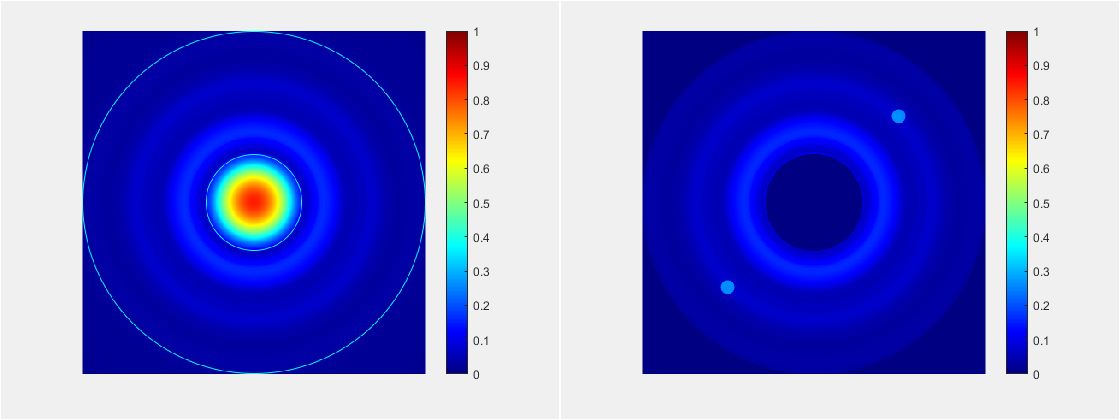}
\mycaption{Source (left) and ground-truth (right) images (original scale).}{The $401 \times 401$ pixel source is a \acs{CC0}-licensed image, \texttt{Airy\_disk\_D65} \cite{Wikipedia2022}, after conversion to grayscale and division by 255. The ground-truth image, shown after pixel values are squared, is the entrywise square root of the source image, after masking (occulting) symmetric inner and outer regions, defined by two circles (left overlays), and after adding spots to simulate twin exoplanets.}
\label{fig:groundtruth}
\end{figure*}

Figure~\ref{fig:groundtruth} introduces a model-based example. It displays in false color, after each pixel value is squared, a \emph{ground-truth} image synthesized from the grayscale version of a \ac{CC0}-licensed \emph{source} image, \texttt{Airy\_disk\_D65}. The latter simulates a focused spot of white light passing through a circular aperture and lens. Rings in the focal or image plane arise due to diffraction. The figure includes, with circles overlaid to define an annular mask, the source image in false color. Using the mask to model a coronagraph, a bright spot is blacked out, representing an occulted star. An outer symmetric region, where diffraction rings are faint, is likewise masked.

Given coronagraphy, the example uses an entrywise square root to make the ground-truth from the source, after conversion from class \texttt{uint8} to \texttt{double} and division by 255. Both images have $401 \times 401$ pixels. Filled circles introduced into the annular foreground simulate twin exoplanets in the ground-truth. Using bicubic and nearest-neighbour interpolation, respectively, ground-truth and mask images are downsampled or upsampled as needed, to vary dimension sizes equally for simulation purposes. The mask is of class \texttt{logical}.

The ground-truth image does not exhibit a phase aberration. To synthesize an \emph{aberrated} image, a random phase offset is introduced entrywise in the Fourier domain, often called the pupil plane in the context of coronagraphy. Because pixel values in the image plane, the inverse Fourier domain, must be real, this phase aberration is constrained to obey the required symmetry. Notwithstanding a few entries, such as the DC value, each entry of the random offset, an image-like matrix, belongs to a uniform distribution of $-\pi$ to $+\pi$ radians.

\begin{figure*}[t]
\centering
\insertpng{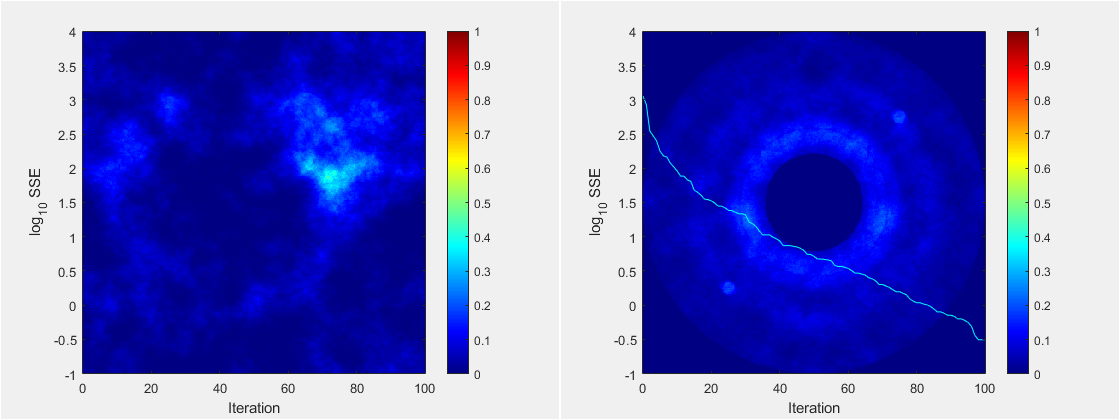}
\mycaption{Input (left) and output (right) images of an enhancement.}{A model-based approach addresses an unknown phase aberration in the pupil plane (Fourier domain) by minimizing an image-plane \acs{SSE}. Nonzero background, and negative foreground, pixels define the \acs{SSE} (right overlay). Diffraction rings and twin exoplanets become visible in the annular foreground when the 2D phase aberration is adequately corrected. Before display, pixel values are squared.}
\label{fig:coronagraph}
\end{figure*}

Figure~\ref{fig:coronagraph} addresses the main part, image enhancement, of the model-based example. It shows the input aberrated image and the output \emph{corrected} image. Prior to display, each pixel value is squared. Starting with an initial guess of the phase aberration, a MATLAB optimization yields a corrected image, approaching the ground-truth, through iteration. The procedure updates the 2D aberration by minimizing a scalar \ac{SSE}, which is overlaid. The \ac{SSE} aggregates deviations in the image plane alone. A corrected image has nonnegative values in the annular foreground and zero values in the background.

The initial guess of the phase aberration is a zero matrix. With it, the optimization converges toward a local minimum at which, although the phase aberration differs (not shown) from the initial random one, the problem is adequately solved, at least in terms of diffraction ring and exoplanet visibility.

In practice, a multistep computing problem is solveable only if it requires a reasonable amount of processing time, measurable in seconds ($\s$), and a reasonable amount of memory space, measurable in bytes ($\B$) of \ac{RAM}, for all steps. With desktop computing, reasonable space means the \ac{RAM} of a \ac{PC} suffices for the image resolution, $401 \times 401$ pixels, at hand. Reasonable time means a fractional second per step suffices on the \ac{PC} at hand, a Dell Latitude E7450 with $8\,\GB$ of \ac{RAM}, for the same.

An accurate and efficient \ac{SSE} software function is required for dimension sizes of interest, given a phase aberration, plus the same for first and second \ac{SSE} derivatives with respect to the 2D aberration. Otherwise, the optimization may fail, due to numerical errors in approximations of the first derivative, the gradient, or may not succeed in reasonable time, due to higher-order curvature not captured without a second derivative, the Hessian, or second-derivative information, a \ac{HMF}. The \ac{RT} algebra enables accurate and efficient \ac{SSE}, gradient, and \ac{HMF} functions in MATLAB.

\begin{figure}[t]
\centering
\insertpdf{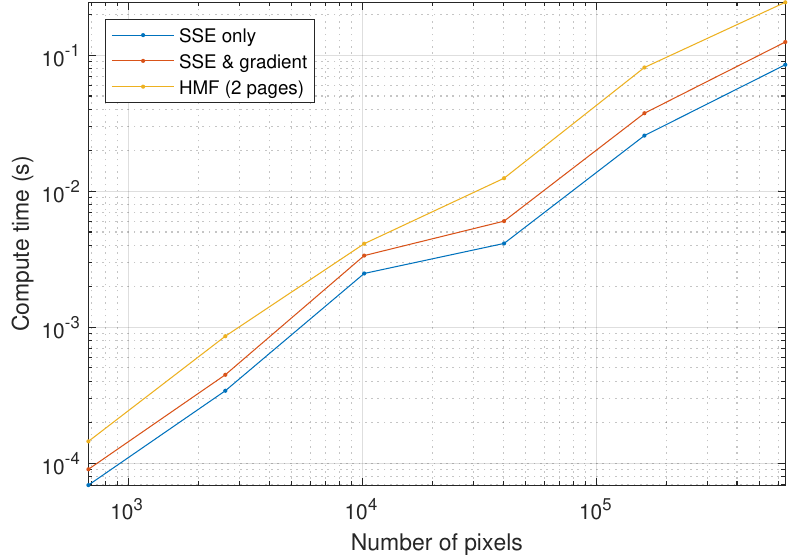}
\mycaption{Compute time for imaging example (multiple scales).}{For an $M \times N$ problem, computing first (gradient) and second (\acs{HMF}) derivatives of the scalar \acs{SSE}, with respect to an $M \times N$ matrix (phase aberration in Fourier domain), is asymptotically equivalent to computing the \acs{SSE} alone, thanks to time efficiencies of a model-based solution derived with the \acs{RT} algebra.}
\label{fig:secsvspels}
\end{figure}

\begin{figure}[t]
\centering
\insertpdf{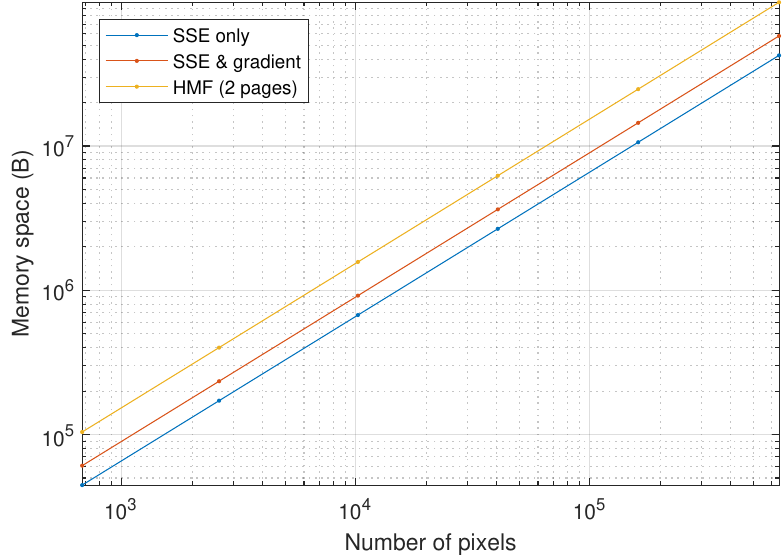}
\mycaption{Memory space for imaging example (multiple scales).}{For an $M \times N$ problem, computing first (gradient) and second (\acs{HMF}) derivatives of the scalar \acs{SSE}, with respect to an $M \times N$ matrix (phase aberration in Fourier domain), is asymptotically equivalent to computing the \acs{SSE} alone, thanks to space efficiencies of a model-based solution derived with the \acs{RT} algebra.}
\label{fig:bytesvspels}
\end{figure}

Figures~\ref{fig:secsvspels} and~\ref{fig:bytesvspels} report the compute time and memory space required, versus the number of pixels, by functions written to determine the \ac{SSE} only, the \ac{SSE} and the gradient, or the \ac{HMF} only. Compute time, averaged over multiple invocations with a \texttt{for} loop, is measured using kernel \texttt{tic} and \texttt{toc} functions. Using kernel \texttt{whos} and \texttt{sum} functions, invoked with negligible overhead, the same MATLAB implementation totals and returns memory space reused by the written functions.

In MATLAB, the \ac{SSE}, the gradient, and the \ac{HMF} may be represented, without the RTToolbox, by a scalar, a 2D array, and a 3D array. Like the gradient, the \ac{HMF} is not computed with every \ac{SSE} computation. When computed, \ac{HMF} time-and-space requirements per page are multiplied by the number of pages. This number, automatically chosen by an Optimization Toolbox routine, \texttt{fminunc}, appeared to be small enough, about two, for the \ac{SSE} to be suitably minimized, as shown in Figure~\ref{fig:coronagraph}, in reasonable time and space overall.

\subsection{Scalar Error}
\label{sec:scalar error}

The image enhancement input is an $M \times N$ aberrated image, $\mathbf{X}^\text{a}$. One defines a scalar \ac{SSE} function in the image plane, an inverse Fourier domain, with a model. Corresponding to an \ac{SSE} minimum, the image enhancement output is an $M \times N$ corrected image, $\mathbf{X}^\text{t}$, an estimate of the ground-truth. Figure~\ref{fig:coronagraph} shows the images, $\mathbf{X}^\text{a}$ and $\mathbf{X}^\text{t}$, after entrywise squares.

To formulate the corrected image, initially unknown, one models an optical system having an entrywise $M \times N$ phase aberration, $\mathbf{\Phi}$, in the pupil plane, a Fourier domain:
\begin{align}
\label{eq:Ya}
\mathbf{Y}^\text{a} &= \mathbf{U} \mathbf{X}^\text{a} \mathbf{V}^\T \text{,} \\
\label{eq:Yt}
\mathbf{Y}^\text{t} &= \mathbf{Y}^\text{a} \circ \exp(\jmath \mathbf{\Phi}) \text{,} \\
\mathbf{X}^\text{t} &= \frac{1}{M N} \bar{\mathbf{U}}^\T \mathbf{Y}^\text{t} \bar{\mathbf{V}} \text{.}
\end{align}
Thus, given an aberration, $\mathbf{\Phi}$, the corrected image's 2D \ac{DFT}, $\mathbf{Y}^\text{t}$, is found from the 2D \ac{DFT}, $\mathbf{Y}^\text{a}$, of the aberrated image, $\mathbf{X}^\text{a}$. An inverse 2D \ac{DFT} yields a corrected image, $\mathbf{X}^\text{t}$.

The 1D \ac{DFT} operators, $\mathbf{U}$ and $\mathbf{V}$, model an (inverse) 2D \ac{DFT} via matrix multiplication. For an $M \times N$ matrix, $\mathbf{X}$, in the image plane and its $M \times N$ counterpart, $\mathbf{Y}$, in the pupil plane, the operators are matrices having dimension sizes $M \times M$ and $N \times N$, respectively. Each entry of the first operator, $\mathbf{U}$, is a complex number whose exponent varies with row and column positions of an integer outer product, $\mathbf{k} \mathbf{k}^\T$, modulo $M$:
\begin{align}
\mathbf{U} &= \exp(-\jmath 2 \pi \mathbf{K} / M) \text{,} \\
\mathbf{K} &= \mathbf{k} \mathbf{k}^\T \mod M \text{,} \\
\mathbf{k}^\T &= \bracket{0 & \dots & M - 1} \text{.}
\end{align}
The second operator, $\mathbf{V}$, is defined similarly. The conjugate transpose of an operator gives its inverse, where a dimension size, $M$ or $N$, provides the scale needed for an identity:
\begin{align}
\frac{1}{M} \bar{\mathbf{U}}^\T \mathbf{U} &= \textbf{I} \text{.}
\end{align}

By the problem definition, the aberrated image, $\mathbf{X}^\text{a}$, is real, and so is the corrected image, $\mathbf{X}^\text{t}$. Thus, the complex-valued 2D \ac{DFT}, $\mathbf{Y}^\text{a}$, exhibits a phase symmetry. The 2D \ac{DFT} of the corrected image, $\mathbf{Y}^\text{t}$, differs only in phase due to an offset. Therefore, for the corrected image to be real, there is an implied symmetry contraint on the phase aberration, $\mathbf{\Phi}$. Discarding the imaginary part, after an inverse 2D \ac{DFT}, deals also with round-off error issues that arise in practice:
\begin{align}
\label{eq:Xt}
\mathbf{X}^\text{t} &= \frac{1}{M N} \real\{\bar{\mathbf{U}}^\T \mathbf{Y}^\text{t} \bar{\mathbf{V}}\} \text{.}
\end{align}

To correct the aberrated image, one requires \emph{a priori} information, in this case \emph{occultation}. The corrected image, when occulted, defines background and foreground errors, $\mathbf{X}^\text{b}$ and $\mathbf{X}^\text{f}$, that are known to be zero in the ground-truth case:
\begin{align}
\mathbf{X}^\text{b} &= \mathbf{W}^\text{b} \circ \mathbf{X}^\text{t} \text{,} \\
\mathbf{X}^\text{f} &= \mathbf{W}^\text{f} \circ \mathbf{X}^\text{t} \text{,} \\
\label{eq:Wf}
\mathbf{W}^\text{f} &= (\mathbf{1} - \mathbf{W}^\text{b}) \circ u(-\mathbf{X}^\text{t}) \text{.}
\end{align}
The occultation mask, $\mathbf{W}^\text{b}$, is an input of the image enhancement. For the example, it is zero at each pixel inside an annulus and one outside it. Note the entrywise product model.

Background and foreground errors test for deviant pixel values, where entries of background and foreground masks, $\mathbf{W}^\text{b}$ and $\mathbf{W}^\text{f}$, are one, respectively. The masks are not complementary because a step function, $u$, filters out nonnegative values in the foreground. However, as the masks are orthogonal, they may be added to help formulate a single error image:
\begin{align}
\label{eq:Xe}
\mathbf{X}^\text{e} &= (\mathbf{W}^\text{b} + \mathbf{W}^\text{f}) \circ \mathbf{X}^\text{t} \text{.}
\end{align}
If the error image, $\mathbf{X}^\text{e}$, deviates from a zero matrix then an aberration, $\mathbf{\Phi}$, requires correction. Thus, the \ac{SSE} is defined as the inner product, $E$, of the error image with itself:
\begin{align}
\label{eq:SSE}
E &= \bar{\mathbf{X}}^\text{e} \bullet \mathbf{X}^\text{e} \text{.}
\end{align}

\subsection{Gradient Matrix}
\label{sec:gradient matrix}

At a minimum of the \ac{SSE}, its gradient, $\mathbf{\nabla}E$, with respect to the phase aberration, $\mathbf{\Phi}$, vanishes. To avoid differentiating a matrix, $\mathbf{X}^\text{e}$, with respect to a matrix, the model employs a degree-two scalar equivalent, $\phi$, of the phase aberration:
\begin{align}
\mathbf{\nabla}E &= (g^{ij} + \bar{g}^{ij}) \mathbf{e}_i \mathbf{e}_j^\T \text{,} \\
g_{ij} &= \frac{\partial\bar{\mathbf{X}}^\text{e}}{\partial\phi^{ij}} \bullet \mathbf{X}^\text{e} \text{,} \\
\mathbf{\Phi} &= \phi^{ij} \mathbf{e}_i \mathbf{e}_j^\T \text{.}
\end{align}
Degree-one basis operators, $\mathbf{e}$ and $\mathbf{e}^\T$, are column and row vectors, each with one nonzero, equal to one, whose vector index corresponds to the tensor index. When the error image, $\mathbf{X}^\text{e}$, is zeroed, the gradient of the \ac{SSE} is also zeroed.

The gradient matrix, $\mathbf{\nabla}E$, appears to depend on an infinite-valued Dirac delta function, $\delta$, the derivative of a Heaviside step function, $u$, which is discontinuous at the origin. First, the partial derivative of the error image, $\mathbf{X}^\text{e}$, is expanded:
\begin{align}
\frac{\partial\mathbf{X}^\text{e}}{\partial\phi^{ij}} &= \frac{\partial\mathbf{W}}{\partial\phi^{ij}} \circ \mathbf{X}^\text{t} + \mathbf{W} \circ \frac{\partial\mathbf{X}^\text{t}}{\partial\phi^{ij}} \text{,} \\
\label{eq:W}
\mathbf{W} &= \mathbf{W}^\text{b} + (\mathbf{1} - \mathbf{W}^\text{b}) \circ u(-\mathbf{X}^\text{t}) \text{,}
\end{align}
where the sum, $\mathbf{W}$, of background and foreground masks, $\mathbf{W}^\text{b}$ and $\mathbf{W}^\text{f}$, is given a symbol. A problematic partial derivative term, which could be infinite valued, is then expanded:
\begin{align}
\frac{\partial\mathbf{W}}{\partial\phi^{ij}} \circ \mathbf{X}^\text{t} &= (\mathbf{1} - \mathbf{W}^\text{b}) \circ \frac{\partial u(-\mathbf{X}^\text{t})}{\partial\phi^{ij}} \circ \mathbf{X}^\text{t} \text{,} \\
\frac{\partial u(-\mathbf{X}^\text{t})}{\partial\phi^{ij}} &= \delta(-\mathbf{X}^\text{t}) \circ -\frac{\partial\mathbf{X}^\text{t}}{\partial\phi^{ij}} \text{.}
\end{align}
Regrouping factors of the expanded derivative, via association and commutation properties of entrywise matrix products, an identity applies. It ensures a zero problematic derivative, via even and sifting properties of the Dirac delta function:
\begin{align}
\delta(\pm\mathbf{X}^\text{t}) \circ \mathbf{X}^\text{t} &= \mathbf{0} \text{.} 
\end{align}

After eliminating the problematic derivative, a degree-two scalar, $g$, that defines the gradient, $\mathbf{\nabla}E$, is rewritten, thanks to an inner, entrywise, and outer product identity and the result, $\mathbf{W}$, of an implied entrywise mask product, $\mathbf{W} \circ \mathbf{W}$:
\begin{align}
g_{ij} &= \left( \mathbf{W} \circ \frac{\partial\bar{\mathbf{X}}^\text{t}}{\partial\phi^{ij}} \right) \bullet \mathbf{X}^\text{e} \text{,} \\
&= \frac{\partial\bar{\mathbf{X}}^\text{t}}{\partial\phi^{ij}} \bullet (\mathbf{W} \circ \mathbf{X}^\text{e}) \text{,} \\
&= \frac{\partial\bar{\mathbf{X}}^\text{t}}{\partial\phi^{ij}} \bullet \mathbf{X}^\text{e} \text{.}
\end{align}
Required for simplification, the inner, entrywise, and outer product identity is derived by considering an equivalent scalar expression that facilitates commutation and association:
\begin{align}
g_{ij} &= \left( w_{k\ell} \frac{\partial\bar{x}_{k\ell}^\text{t}}{\partial\phi^{ij}} \right) x_\text{e}^{k\ell} \text{,} \\
&= \frac{\partial\bar{x}_{k\ell}^\text{t}}{\partial\phi^{ij}} (w^{k\ell} x_\text{e}^{k\ell}) \text{.}
\end{align}

Subsequently, using the inverse 2D \ac{DFT} model, the partial derivative of the corrected image, $\mathbf{X}^\text{t}$, with respect to the degree-two scalar aberration, $\phi$, is expanded as follows:
\begin{align}
\frac{\partial\mathbf{X}^\text{t}}{\partial\phi^{ij}} &= \frac{1}{M N} \bar{\mathbf{U}}^\T \frac{\partial\mathbf{Y}^\text{t}}{\partial\phi^{ij}} \bar{\mathbf{V}} \text{,} \\
\frac{\partial\mathbf{Y}^\text{t}}{\partial\phi^{ij}} &=  \mathbf{Y}^\text{t} \circ \jmath \frac{\partial\mathbf{\Phi}^\text{t}}{\partial\phi^{ij}} \text{,} \\
\frac{\partial\mathbf{\Phi}^\text{t}}{\partial\phi^{ij}} &= \mathbf{e}_i \mathbf{e}_j^\T \text{.}
\end{align}

Similar to the nonconjugate case, the partial derivative of the corrected image's conjugate, $\bar{\mathbf{X}}^\text{t}$, with respect to the degree-two scalar aberration, $\phi$, may be expressed as follows:
\begin{align}
\frac{\partial\bar{\mathbf{X}}^\text{t}}{\partial\phi^{ij}} &= \frac{-\jmath}{M N} \mathbf{U}^\T (\bar{\mathbf{Y}}^\text{t} \circ \mathbf{e}_i \mathbf{e}_j^\T) \mathbf{V} \text{.}
\end{align}
To simplify this expression, as follows, the enclosed matrix, $\mathbf{Y}^\text{t}$, converts to an equivalent degree-two scalar, $y^\text{t}$:
\begin{align}
\mathbf{Y}^\text{t} \circ \mathbf{e}_i \mathbf{e}_j^\T &= y_{ij}^\text{t} \mathbf{e}_i \mathbf{e}_j^\T \text{.}
\end{align}
As a scalar, of any degree, may be factored out of a multi-operand product, the partial derivative simplifies:
\begin{align}
\frac{\partial\bar{\mathbf{X}}^\text{t}}{\partial\phi^{ij}} &= \frac{-\jmath}{M N} \bar{y}_{ij}^\text{t} \mathbf{U}^\T \mathbf{e}_i \mathbf{e}_j^\T \mathbf{V} \text{.}
\end{align}

Thus, the degree-two scalar, $g$, may be expressed in terms of another degree-two scalar, denoted $y^\text{e}$, as follows:
\begin{align}
g_{ij} &= \frac{-\jmath}{M N} \bar{y}_{ij}^\text{t} y_{ij}^\text{e} \text{,} \\
y_{ij}^\text{e} &= (\mathbf{U}^\T \mathbf{e}_i \mathbf{e}_j^\T \mathbf{V}) \bullet \mathbf{X}^\text{e} \text{.}
\end{align}
By replacing the basis operators, $\mathbf{e}$ and $\mathbf{e}^\T$, the 1D \ac{DFT} operators, $\mathbf{U}$ and $\mathbf{V}$, and the error image, $\mathbf{X}^\text{e}$, with equivalent degree-two scalars, including Dirac symbols, $\delta$, the following reformulation arises, thanks to the \ac{RT} algebra:
\begin{align}
y_{ij}^\text{e} &= (u_{\alpha k} \delta_i^\alpha \delta_j^\beta v_{\beta\ell}) x_\text{e}^{k\ell} \text{,} \\
&= u_{ik} x_\text{e}^{k\ell} v_{j\ell} \text{.}
\end{align}
If each degree-two scalar, e.g., $x_\text{e}$, is now replaced with its equivalent matrix, i.e., $\mathbf{X}^\text{e}$, a hidden 2D \ac{DFT} reveals:
\begin{align}
\mathbf{G} &= \frac{-\jmath}{M N} \bar{\mathbf{Y}}^\text{t} \circ \mathbf{Y}^\text{e} \text{,} \\
\label{eq:Ye}
\mathbf{Y}^\text{e} &= \mathbf{U} \mathbf{X}^\text{e} \mathbf{V}^\T \text{.}
\end{align}

Consequently, the matrix equivalent, $\mathbf{G}$, of the degree-two scalar, $g$, is directly proportional to an entrywise product, in the Fourier domain with one operand conjugated, of the corrected image, $\mathbf{X}^\text{t}$, and the error image, $\mathbf{X}^\text{e}$. As the gradient matrix, $\mathbf{\nabla}E$, equals twice the real part, $\real\{\mathbf{G}\}$, of this equivalent, its final formulation proves to be relatively simple:
\begin{align}
\mathbf{\nabla}E &= \mathbf{G} + \bar{\mathbf{G}} \text{,} \\
\label{eq:GSSE}
&= \frac{2}{M N} \imag\{\bar{\mathbf{Y}}^\text{t} \circ \mathbf{Y}^\text{e}\} \text{.}
\end{align}

\subsection{Hessian Function}
\label{sec:hessian function}

For the phase aberration correction problem, once the \ac{SSE} and its gradient are formulated, one may attempt to construct a solution using an established routine, like \texttt{fminunc} from MATLAB's Optimization Toolbox. This function offers the user a choice between quasi-Newton and trust-region algorithms. Because the former proves too slow even on small-format test cases, all simulations employ the latter. With the quasi-Newton algorithm, the gradient is actually optional. With the trust-region algorithm, the gradient is required.

Asymptotically, it takes equal time and space to construct the \ac{SSE} and gradient as the \ac{SSE} alone. Using explicit \ac{DFT} operators, either takes $O(M^2 N + M N^2)$ time and $O(M N)$ space. Each \ac{DFT} operator represents a Fourier transform, or its inverse, as matrix multiplication. One need not construct the operator, a useful formalism, as the \ac{FFT} constructs a \ac{DFT} faster. Using 2D \acp{FFT}, the \ac{SSE} and gradient or the \ac{SSE} only takes $O(M N \log M N)$ time and $O(M N)$ space.

Non-asymptotically, it takes longer to compute the \ac{SSE} and gradient than it does to compute the \ac{SSE} alone. Yet without the gradient, on small-format test cases, the trust-region algorithm was too slow. For medium and large-format test cases, even the benefit of an explicit first-order partial derivative, the gradient, was insufficient. For a practical solution, an implicit second-order partial derivative, an \ac{HMF}, was needed. The \ac{RT} algebra helps to formulate and simplify the desired \ac{HMF}.

On one hand, the Hessian, $\mathbf{H}$, of the scalar \ac{SSE}, $E$, is a degree-two matrix that requires at least $O(M^2 N^2)$ space and, therefore, time to construct, even with \ac{FFT} acceleration:
\begin{align}
\mathbf{H}_{ij} &= \frac{\partial\mathbf{\nabla}E}{\partial\phi^{ij}} \text{.}
\end{align}
On the other hand, the \ac{HMF}, $\mathbf{F}$, is a degree-one matrix that requires at least $O(M N P)$ space and time to construct, where one dimension size, $P$, refers to a third index, $k$, of a phase step, $\Delta\phi$, a degree-three, not a degree-two, scalar here:
\begin{align}
\mathbf{F}_k &= \mathbf{H}_{ij} \Delta\phi_k^{ij} \text{.}
\end{align}

According to details of the trust-region algorithm, the \ac{HMF} is computed occasionally only after a gradient computation. Referring to 2D \acp{DFT}, $\mathbf{Y}^\text{t}$ and $\mathbf{Y}^\text{e}$, of corrected and error images, $\mathbf{X}^\text{t}$ and $\mathbf{X}^\text{e}$, the \ac{HMF} is readily expanded:
\begin{align}
\mathbf{F}_k &= \frac{2}{M N} \imag\left\{ \frac{\partial\bar{\mathbf{Y}}^\text{t}}{\partial\phi^{ij}} \circ \mathbf{Y}^\text{e} + \bar{\mathbf{Y}}^\text{t} \circ \frac{\partial\mathbf{Y}^\text{e}}{\partial\phi^{ij}} \right\} \Delta\phi_k^{ij} \text{,} \\
\label{eq:HMF}
&= \frac{2}{M N} \imag\{\Delta\bar{\mathbf{Y}}_k^\text{t} \circ \mathbf{Y}^\text{e} + \bar{\mathbf{Y}}^\text{t} \circ \Delta\mathbf{Y}_k^\text{e}\} \text{,}
\end{align}
where \ac{HMF} components, $\Delta\bar{\mathbf{Y}}^\text{t}$ and $\Delta\mathbf{Y}^\text{e}$, represent a sequence of 2D \ac{DFT} increments, with one of them conjugated. The first \ac{HMF} component, $\Delta\mathbf{Y}^\text{t}$, is also readily formulated:
\begin{align}
\Delta\mathbf{Y}_k^\text{t} &= (\mathbf{Y}^\text{t} \circ \jmath \mathbf{e}_i \mathbf{e}_j^\T) \Delta\phi_k^{ij} \text{,} \\
\label{eq:DYt}
&= \jmath \mathbf{Y}^\text{t} \circ \Delta\mathbf{\Phi}_k \text{,} \\
\Delta\mathbf{\Phi}_k &= \Delta\phi_k^{ij} \mathbf{e}_i \mathbf{e}_j^\T \text{,}
\end{align}
where the degree-three scalar, $\Delta\phi$, is replaced with an equivalent degree-one matrix, $\Delta\mathbf{\Phi}$. The incremental time and space cost of constructing this conversion, if it were actually needed beyond a formalism, can be made practically zero.

The second \ac{HMF} component, $\Delta\mathbf{Y}^\text{e}$, is partially formulated by revealing a sequence of 2D \acp{DFT} as follows:
\begin{align}
\Delta\mathbf{Y}_k^\text{e} &= \left( \mathbf{U} \frac{\partial\mathbf{X}^\text{e}}{\partial\phi^{ij}} \mathbf{V}^\T \right) \Delta\phi_k^{ij} \text{,} \\
\label{eq:DYe}
&= \mathbf{U} \Delta\mathbf{X}_k^\text{e} \mathbf{V}^\T \text{,} \\
\Delta\mathbf{X}_k^\text{e} &= \left( \mathbf{W} \circ \frac{\partial\mathbf{X}^\text{t}}{\partial\phi^{ij}} \right) \Delta\phi_k^{ij} \text{,} \\
\label{eq:DXe}
&= \mathbf{W} \circ \Delta\mathbf{X}_k^\text{t} \text{.}
\end{align}
To complete the formulation, a sequence of inverse 2D \acp{DFT} of the first \ac{HMF} component, $\Delta\mathbf{Y}_k^\text{t}$, is revealed:
\begin{align}
\Delta\mathbf{X}_k^\text{t} &= \frac{\jmath}{M N} \bar{\mathbf{U}}^\T (\mathbf{Y}^\text{t} \circ \mathbf{e}_i \mathbf{e}_j^\T) \bar{\mathbf{V}} \,\Delta\phi_k^{ij} \text{,} \\
&= \frac{\jmath}{M N} \bar{\mathbf{U}}^\T (\mathbf{Y}^\text{t} \circ \Delta\mathbf{\Phi}_k) \bar{\mathbf{V}} \text{,} \\
\label{eq:DXt}
&= \frac{1}{M N} \bar{\mathbf{U}}^\T \Delta\mathbf{Y}_k^\text{t} \bar{\mathbf{V}} \text{.}
\end{align}

Thanks to the \ac{RT} algebra, the \ac{HMF} asymptotically requires, with 2D \acp{FFT}, $O(M N P \log M N)$ time and $O(M N P)$ space to compute. This asymptotic complexity is directly proportional by a factor, $P$, to that of the gradient. The incremental time-and-space complexity of computing the \ac{HMF}, $\mathbf{F}$, with the gradient, $\mathbf{\nabla}E$, is $O(1)$ asymptotically if $P$, which depends on the optimization routine, \texttt{fminunc}, is bounded above.

\subsection{Implementation}
\label{sec:implementation}

Figures~\ref{fig:ssefun} and~\ref{fig:hessmfun} summarize the phase aberration correction solution. These figures give MATLAB functions to compute the \ac{SSE}, its gradient, and its \ac{HMF}, with software statements cross-referenced to algebraic equations. The code also shows how to compute a corrected image, $\mathbf{X}^\text{t}$, from the aberrated image, $\mathbf{X}^\text{a}$, and an optimizable phase aberration, $\mathbf{\Phi}$.

\begin{figure}[t]
\centering
\begin{tabular}{l}
\hline
\texttt{function [E,GrE,Xt] = ssefun(Ph,Xa,Wb)} \\
\hline
\texttt{Ya = fft2(Xa); \% \eqref{eq:Ya}} \\
\texttt{Yt = Ya.*complex(cos(Ph),sin(Ph)); \% \eqref{eq:Yt}} \\
\texttt{Xt = real(ifft2(Yt)); \% \eqref{eq:Xt}} \\
\texttt{Wf = \~{}Wb \& Xt < 0; \% \eqref{eq:Wf}} \\
\texttt{Xe = (Wb | Wf).*Xt; \% \eqref{eq:Xe}} \\
\texttt{if nargout > 1} \\
\texttt{~~~~MN = numel(Xt);} \\
\texttt{~~~~Ye = fft2(Xe); \% \eqref{eq:Ye}} \\
\texttt{~~~~GrE = 2/MN*imag(conj(Yt).*Ye); \% \eqref{eq:GSSE}} \\
\texttt{end} \\
\texttt{Xe = Xe(:);} \\
\texttt{E = Xe'*Xe; \% \eqref{eq:SSE}} \\
\hline
\end{tabular}
\mycaption{Model-based \acs{SSE} and gradient.}{Developed for the imaging example, this MATLAB function's output arguments are the \acs{SSE}, $E$, the \acs{SSE} gradient, $\mathbf{\nabla}E$, and a corrected image, $\mathbf{X}^\text{t}$. Its input arguments are a phase aberration, $\mathbf{\Phi}$, the aberrated image, $\mathbf{X}^\text{a}$, and the occultation (background) mask, $\mathbf{W}^\text{b}$.}
\label{fig:ssefun}
\end{figure}

\begin{figure}[t]
\centering
\begin{tabular}{l}
\hline
\texttt{function F = hessmfun(Xt,DPh,Wb)} \\
\hline
\texttt{Yt = fft2(Xt); \% \eqref{eq:Yt}} \\
\texttt{W = Wb | (\~{}Wb \& Xt < 0); \% \eqref{eq:W}} \\
\texttt{Ye = fft2(W.*Xt); \% \eqref{eq:Xe} \& \eqref{eq:Ye}} \\
\texttt{[M,N] = size(Xt);} \\
\texttt{DPh = reshape(DPh,M,N,[]);} \\
\texttt{MN = numel(Xt);} \\
\texttt{DYt = 1j*(Yt.*DPh); \% \eqref{eq:DYt}} \\
\texttt{DXt = real(ifft2(DYt)); \% \eqref{eq:DXt}} \\
\texttt{DYe = fft2(W.*DXt); \% \eqref{eq:DXe} \& \eqref{eq:DYe}} \\
\texttt{DYY = imag(conj(DYt).*Ye); \% \eqref{eq:HMF}} \\
\texttt{YDY = imag(conj(Yt).*DYe); \% \eqref{eq:HMF}} \\
\texttt{F = 2/MN*(DYY+YDY); \% \eqref{eq:HMF}} \\
\texttt{F = reshape(F,MN,P);} \\
\hline
\end{tabular}
\mycaption{Model-based multipage \acs{HMF}.}{Invoked occasionally after another function, \texttt{ssefun}, this MATLAB function's output argument is the \acs{HMF}, $\mathbf{F}$. Its input arguments are a corrected image, $\mathbf{X}^\text{t}$, a multipage phase aberration step, $\Delta\mathbf{\Phi}$, and the occulatation mask, $\mathbf{W}^\text{b}$. Some lines use broadcasting.}
\label{fig:hessmfun}
\end{figure}

The MATLAB functions, \texttt{ssefun} and \texttt{hessmfun}, do not require the RTToolbox. However, they exploit efficiencies derived using the \ac{RT} algebra. Figures~\ref{fig:ssefun} and~\ref{fig:hessmfun} clarify the (inverse) 2D \acp{DFT} revealed after algebraic manipulations that involved purely numeric aspects of tensor operands, some nonzero degree. The dual-variant index notation excelled on a variety of inner, entrywise, outer, and mixed products.

Where one of two operands is multipage, a 3D array, the \ac{HMF}, \texttt{hessmfun}, leverages \ac{EMV} broadcasting. An operator, \texttt{.*}, enables entrywise products with pagewise broadcasting. The \ac{HMF} uses pagewise kernel routines, \texttt{fft2} and \texttt{ifft2}, also. Thus, the example motivates nonscalar and broadcasting extensions of Ricci notation featured by the \ac{RT} algebra.

%%%%%%%%%%%%%%%%%%%%%%%%%%%%%%%%%%
% Tensor Software
%%%%%%%%%%%%%%%%%%%%%%%%%%%%%%%%%%

\section{Tensor Software}
\label{sec:tensor software}

Table~\ref{tab:rttoolbox} introduces the RTToolbox. Together with the classes and functions of the MATLAB kernel, compiled or interpreted, the RTToolbox defines the \ac{RT} software. First, this section offers a design overview of \texttt{tensor} objects. Next, it presents \texttt{index} objects, which play a critical role in parsing a dual-variant index notation. Finally, key problems and solutions, contrasted with closest equivalents of the \ac{NT} software, are summarized for unary, binary, and $N$-ary operations.

\begin{table}[t]
\centering
\mycaption{Composition of the RTToolbox.}{The \acs{RT} software is defined as the RTToolbox, comprising \texttt{tensor} and \texttt{index} classes, pagewise functions, and unit tests, and the MATLAB kernel. Together, they naturally express and compute the \ac{RT} algebra. There are no dependencies on other toolboxes (or libraries).}
\label{tab:rttoolbox}
\vspace{1ex}
\begin{tabular}{lll}
\hline
M-file & Summary & Note \\
\hline
\texttt{tensor.m} & Definition of \texttt{tensor} objects & Table~\ref{tab:tensorclass} \\
\texttt{tensorTest.m} & Tests of \texttt{tensor} objects & Table~\ref{tab:tensortest} \\
\texttt{index.m} & Definition of \texttt{index} objects & Table~\ref{tab:indexclass} \\
\texttt{indexTest.m} & Tests of \texttt{index} objects & Table~\ref{tab:indextest} \\
\hline
\texttt{pagetrace.m} & Pagewise \texttt{trace} function & Figure~\ref{fig:secsvspages} \\
\texttt{pagediag.m} & Pagewise \texttt{diag} function & - \\
\texttt{pagehorzcat.m} & Pagewise \texttt{horzcat} function & Figure~\ref{fig:secsvspages} \\
\texttt{pagevertcat.m} & Pagewise \texttt{vertcat} function & - \\
\texttt{pagecat.m} & Pagewise \texttt{cat} function & - \\
\hline
\end{tabular}
\end{table}

\subsection{Tensor Objects}
\label{sec:tensor objects}

Table~\ref{tab:tensorclass} summarizes the \texttt{tensor} class. Apart from the first group, constructor aside, and \texttt{isequal}, all methods return a \texttt{tensor} object. Table~\ref{tab:tensortest} presents excerpts, cross-referenced to selected equations, from unit tests of class \texttt{tensor}.

\begin{table}[t]
\centering
\mycaption{Tensor object class definition.}{Favouring simplicity, this release of the \texttt{tensor} class has a constructor and just these public methods, including overloaded operators. Ordinary methods are grouped to further simplify the discussion.}
\label{tab:tensorclass}
\vspace{1ex}
\begin{tabular}{l}
\hline
Public methods (\texttt{tensor} class) \\
\hline
Constructor, \emph{etc.} \\
\hspace{1em} \verb|tensor| (0, 1, 2+ args in), \verb|index|, \verb|entry|, \verb|degree|, \verb|ndims|, \\
\hspace{1em} \verb|numel|, \verb|length|, \verb|size|, \verb|end| (\verb|(,)|) \\
Unary operations \\
\hspace{1em} \verb|subsref| (\verb|(,)|), \verb|sum|, \verb|permute|, \verb|uminus| (\verb|-|), \verb|uplus| (\verb|+|), \\
\hspace{1em} \verb|conj|, \verb|not| (\verb|~|), \verb|abs|, \verb|log|, \verb|round|, \verb|transpose| (\verb|.'|), \\
\hspace{1em} \verb|ctranspose| (\verb|'|), \verb|trace|, \verb|diag| \\
Binary operations \\
\hspace{1em} \verb|subsasgn| (\verb|(,)=|), \verb|plus| (\verb|+|), \verb|minus| (\verb|-|), \verb|eq| (\verb|==|), \verb|ne| (\verb|~=|), \\
\hspace{1em} \verb|lt| (\verb|<|), \verb|gt| (\verb|>|), \verb|le| (\verb|<=|), \verb|ge| (\verb|>=|), \verb|and| (\verb|&|), \verb|or| (\texttt{|}), \\
\hspace{1em} \verb|times| (\verb|.*|), \verb|ldivide| (\verb|.\|), \verb|rdivide| (\verb|./|), \verb|power| (\verb|.^|), \\
\hspace{1em} \verb|mtimes| (\verb|*|), \verb|mldivide| (\verb|\|), \verb|mrdivide| (\verb|/|) \\
$N$-ary operations \\
\hspace{1em} \verb|horzcat| (\verb|[,]|), \verb|vertcat| (\verb|[;]|), \verb|cat|, \verb|isequal| \\
\hline
\end{tabular}
\end{table}

\begin{table}[t]
\centering
\mycaption{Tensor object unit tests.}{Included with the RTToolbox, the \texttt{tensor} object unit tests were developed for this paper. To illustrate programmatic efficiency, software expressions are cross-referenced to equivalent algebraic expressions. Here, DN means any degree and D\# means degree~\#.}
\label{tab:tensortest}
\vspace{1ex}
\begin{tabular}{ll}
\hline
Algebraic expression (ref.) & Software expression \\
\hline
Inner product, D1 scalar \eqref{eq:inner1} & \verb|a(i)*b(i)*c(~i)| \\
Entrywise product, D1 scalar \eqref{eq:ewise} & \verb|a(i)*b(i)*c(i)| \\
Outer product, D1 scalar \eqref{eq:outer1} & \verb|a(i)*b(j)*c(k)| \\
Entrywise relation, D2 scalar \eqref{eq:relate1} & \verb|y(i,~j) ~= y(~j,i))| \\
Permute and copy, D2 scalar \eqref{eq:permute} & \verb|z(i,~j) = y(~j,i)| \\
Left division, D2 scalar \eqref{eq:divide} & \verb|A(l,lp)\b(i,lp)| \\
Mixed product, D2 scalar \eqref{eq:mixed1} & \verb|A(l,lp)*u(i,~l)| \\
Outer addition, D1 scalar \eqref{eq:outer2} & \verb|log(c(j)+x(i))| \\
\hline
Mixed product, D2 matrix \eqref{eq:mixed2} & \verb|A(i,~j)*B(i,~j)| \\
Trace contraction, D2 matrix \eqref{eq:contract} & \verb|trace(C(i,~i))| \\
Diagonal attraction, D2 matrix \eqref{eq:attract} & \verb|diag(C(i,i))| \\
Inner product, DN vector \eqref{eq:inner2} & \verb|x(~k)'*x(~k)| \\
Col.\ concatenation, D1 vector \eqref{eq:concat} & \verb|[ones(M,1) abs(c(i))]| \\
Row concatenation, D1 vector \eqref{eq:concat} & \verb|[b(j).'; ones(1,N)]| \\
Outer relation, D1 vector \eqref{eq:relate2} & \verb|A(i,~j) >= b(j).'| \\
Index concat., D2 matrix \eqref{eq:indcat} & \verb|cat(j,A(i,j),B(j,k))| \\
\hline
\end{tabular}
\end{table}

A \texttt{tensor} object has two protected properties, \texttt{indices} and \texttt{entries}, which the \texttt{index} and \texttt{entry} methods offer a copy of, respectively. The \texttt{indices} property is a row vector, possibly empty, of class \texttt{index}. The \texttt{entries} property must be a non-\texttt{tensor} array, an \ac{MDA}, usually of class \texttt{double} or \texttt{logical}, the only classes represented in unit tests.

For a \texttt{tensor} object, the \texttt{degree} method returns the number of indices, i.e., the \texttt{numel} of the \texttt{indices} property, and the \texttt{ndims} method returns the number of dimensions, i.e., the \texttt{ndims} of the \texttt{entries} property. Because first and second dimensions are \ac{MDA} rows and columns, the \texttt{degree} is at least two less than the \texttt{ndims} of a \texttt{tensor} object.

The one-argument constructor returns a \texttt{tensor} object, whose \texttt{entries} and \texttt{degree} equal the non-\texttt{tensor} argument and its \texttt{ndims} minus two. A corresponding \texttt{index} object is constructed and assigned to the \texttt{indices}. A zero-argument, or default, constructor returns a \texttt{tensor} object of \texttt{degree} zero with empty \texttt{entries} and \texttt{indices}.

The last constructor form enables a \texttt{tensor} object of specifiable \texttt{indices} and \texttt{degree}, in addition to \texttt{entries}. A first non-\texttt{tensor} argument is assigned to the \texttt{entries}. Additional arguments must be \texttt{index} scalars or row vectors. The concatenated \texttt{index} arguments, extended if the result is too short compared to \texttt{ndims} minus two of the \texttt{entries}, is assigned to the \texttt{indices}. The \texttt{degree} exceeds \texttt{ndims} minus two when \texttt{index} arguments suffice to address trailing singleton (\texttt{size} one) dimensions of the \texttt{entries}.

Additional methods, \texttt{numel} and \texttt{length}, return the product of all dimension sizes and the size of the largest non-singleton dimension, respectively. They equal the \texttt{numel} and \texttt{length} of the \texttt{entries}. The \texttt{size} method invokes the \texttt{size} function, passing along parameters, on the \texttt{entries}. When all parameters are \texttt{index} objects, they convert first to numbers that address \ac{MDA} dimensions. The \texttt{end} method, needed for subscripting operations, \verb|(,)| and \verb|(,)=|, with numeric subscripts, requires \texttt{builtin} to invoke the \texttt{end} function on the \texttt{entries}, as \texttt{end} is also a keyword.

The \ac{RT} software uses \texttt{index} objects with a logical NOT operator to represent \texttt{false}-variant indices as subscripts, alongside \texttt{true}-variant ones. Whereas Table~\ref{tab:tensortest} suggests \texttt{index} subscripts are always required for nonzero-degree \texttt{tensor} expressions, the \ac{RT} software enables a context-dependent \texttt{index}-free notation compatible with the \ac{RT} algebra. For example, the inner product of an any-degree vector, \verb|x|, with itself has a simpler form, \verb|x'*x|, than the one shown. A \texttt{tensor} object always has \texttt{indices}, which may be specified or returned upon construction. Expressions that yield \texttt{tensor} results imply associated \texttt{indices} in predictable ways.

\subsection{Index Objects}
\label{sec:index objects}

Table~\ref{tab:indexclass} summarizes the constructor and ordinary methods, including overloaded operators, of the \texttt{index} class. Table~\ref{tab:indextest} presents selected unit tests of class \texttt{index}. When a \texttt{tensor} object is constructed, its \texttt{indices} may be simultaneously dealt, as one or more \texttt{index} objects, to extra output arguments on the \ac{LHS} of an assignment.

\begin{table}[t]
\centering
\mycaption{Index object class definition.}{Having few of its own methods, the \texttt{index} class inherits copy-by-reference, (in)equality, and concatenation operators from the \texttt{handle} class. Inheritance enables use of set-theoretic functions exploited by the \texttt{tensor} class to parse an extended Ricci notation.}
\label{tab:indexclass}
\vspace{1ex}
\begin{tabular}{l}
\hline
Public methods (\texttt{index} class) \\
\hline
Constructor, \emph{etc.} \\
\hspace{1em} \verb|index| (0, 1 args in), \verb|deal| \\
Unary methods \\
\hspace{1em} \verb|logical|, \verb|true|, \verb|false|, \verb|not| (\verb|~|) \\
\hline
\end{tabular}
\end{table}

\begin{table}[t]
\centering
\mycaption{Unit tests of \texttt{index} objects.}{Each entry of an \texttt{index} object, which may be a vector, expresses a \texttt{true}-variant or \texttt{false}-variant state. These tests illustrate: (1)~a scalar \texttt{index}, (2)~a vector \texttt{index}, (3)~a one-step \texttt{deal}, (4)~a with-\texttt{tensor} construction, and (5)~a cast-to-\texttt{logical}.}
\label{tab:indextest}
\begin{tabular}{l|l}
\verb|%% Test 1| & \verb|%% Test 3| \\
\verb|k = index;| & \verb|[i,j] = index(2);| \\
\verb|assert(isscalar(k))| & \verb|assert(isscalar(i))| \\
\verb|assert(k ~= ~k)| & \verb|assert(isscalar(j))| \\
\verb|assert(k == ~~k)| & \verb|%% Test 4| \\
\verb|assert(true(~k) == k)| & \verb|sz = [100 100 10];| \\
\verb|assert(false(k) == ~k)| & \verb|[A,k] = tensor(rand(sz));| \\
\verb|%% Test 2| & \verb|assert(isa(k,'index'))| \\
\verb|k = index(2);| & \verb|%% Test 5| \\
\verb|[i,j] = deal(k);| & \verb|k = [~index index(2)];| \\
\verb|assert(isscalar(i))| & \verb|var = logical(k);| \\
\verb|assert(isscalar(j))| & \verb|ftt = [false true true];| \\
\verb|assert(all(k == [i j]))| & \verb|assert(isequal(var,ftt))| \\
\end{tabular}
\end{table}

A scalar \texttt{index} is constructed as follows. The object has two protected properties, \texttt{tilde}, itself an \texttt{index} object, and \texttt{state}, a \texttt{logical} variable. The \texttt{tilde} property of the \texttt{index} object's \texttt{tilde} property equals the \texttt{index} object itself. Moreover, the \texttt{state} of the \texttt{index} object and the \texttt{state} of its \texttt{tilde} property are always complementary. Upon construction, they are \texttt{true} and \texttt{false}, respectively. The constructor accomplishes this feat thanks in part to inheritance of the \texttt{handle} class and also via one-level recursion. A \texttt{handle} object resembles a \texttt{void} pointer in C/C++.

For a scalar \texttt{index} object, the \texttt{logical} method returns its \texttt{state} property (class \texttt{logical}) and the \texttt{true} or \texttt{false} methods return the \texttt{index} object itself or its \texttt{tilde} property (class \texttt{index}), whichever one has a \texttt{state} property equal to logical \texttt{true} or \texttt{false}, respectively. The NOT operator, overloaded by the \texttt{not} method, always returns the \texttt{tilde} property. In this way, respectively, the \texttt{true}-variant, \texttt{false}-variant, or complementary \texttt{index} object is obtained. As with the \texttt{logical} method, the \texttt{true}, \texttt{false}, and \texttt{not} methods work entrywise if the \texttt{index} object is nonscalar.

Because it inherits from class \texttt{handle}, class \texttt{index} does not need a \texttt{horzcat} method to enable \texttt{index} object concatenation into a row vector of class \texttt{index}. Kernel functions, like \texttt{length}, also work as expected. A one-argument form of the \texttt{index} constructor accepts a number to construct a row vector of class \texttt{index} with \texttt{length} equal to that number. The public \texttt{deal} method is invoked by the constructor to deal each entry of the row vector to output arguments in sequence. If there are fewer output arguments than the vector \texttt{length}, the last output argument is the vector remainder.

Upon construction, entries of an \texttt{index} object and their complements are unique, due to the \texttt{handle} superclass. Copy assignment, equality, and inequality operators are inherited. Entrywise equality of one \texttt{index} object to another, or to its complement via the entrywise NOT operator, offers a way to identify repeated indices, whether of the same variant or not, in a \texttt{tensor} expression. Entrywise relational operators mean that set-theoretic functions like \texttt{ismember} and \texttt{unique}, which invoke a kernel \texttt{sort}, may be used with the \texttt{index} objects. As they are currently an interpreted part of the kernel, set-theoretic functions incur abstraction penalties.

The \ac{NT} software uses template metaprogramming (LibNT), character vectors (NTToolbox), and counters (LibNT and NTToolbox) to parse the multi-variant indices with underline notation of the \ac{NT} algebra. In contrast, each \texttt{index} object of the \ac{RT} software always has only two variants and, with the \texttt{tensor} class, the \texttt{index} class helps to parse overscript bar notation, which can represent superscript/subscript notation with dot spacers as required, of the \ac{RT} algebra.

\subsection{Unary Operations}
\label{sec:unary operations}

A \texttt{tensor} may be subscripted with \texttt{index}, numeric, or other objects on the \ac{RHS} or \ac{LHS} of an (implied) assignment via overloaded \texttt{subsref} and \texttt{subsasgn} operators, respectively. Enclosing comma-separated subscripts in parentheses, the \texttt{subsref} operator is considered a unary operator on the \texttt{tensor} the parentheses follow. The \texttt{subsasgn} operator is considered a binary operator where one operand is the \texttt{tensor} the parentheses follow, on the \ac{LHS} of the assignment, and the other operand is the result of an expression, on the \ac{RHS} of the assignment.

When the \texttt{subsref} method is invoked, either all or none of the subscripts must be \texttt{index} objects. In the second case, the result of the \texttt{subsref} expression is an equivalent \texttt{subsref} expression applied to the \texttt{entries} of the \texttt{tensor}. The first case invokes a \texttt{simplify} method or throws an error. To avoid the runtime error, there must be at least as many indices, when subscripts are concatenated, as the \texttt{degree} of the \texttt{tensor}. Extra indices address trailing singleton dimensions of the \texttt{entries} and increase the \texttt{degree} of the \texttt{tensor} result. This feature supports a variety of outer operations.

The protected \texttt{simplify} method does two things. First, for each set of indices that match, irrespective of variants, an attraction is performed. Second, for each set of attracted indices, a contraction is performed if at least one member is of the complementary variant. Attracted indices that are not contracted express unique variants that are determined and assigned within the \texttt{indices} property of the returned \texttt{tensor}. All attractions are performed via one invocation of a protected method, \texttt{select}, that leverages kernel support for linear indexing. All contractions are performed via one invocation of an overloaded and public \texttt{sum} method.

After converting \texttt{index} arguments to numeric dimensions, the \texttt{sum} method invokes a kernel \texttt{sum} on the \texttt{entries} of a \texttt{tensor}, passing on arguments. First, a kernel \texttt{ismember} tests membership of concatenated \texttt{index} arguments in the \texttt{indices}, finding dimensions of \texttt{entries}. With the \texttt{true} method, the \texttt{index} test is done irrespective of variants. Before invoking \texttt{sum} on the \texttt{entries}, dimensions from matched \texttt{indices} are increased by two, as first and second dimensions address rows and columns of \texttt{entries}, respectively. Unmatched \texttt{index} arguments address trailing singleton dimensions, which always exist, of the \texttt{entries}.

Though the \texttt{subsasgn} operator is binary, it relies on the unary \texttt{permute} method. The latter requires either all-\texttt{index} or all-numeric arguments. Before invoking a kernel \texttt{permute} on the \texttt{entries} property, passing along all arguments, the \texttt{index} arguments find equivalent numeric dimensions. The \texttt{permute} method updates the sequence of returned indices, which make up the \texttt{indices} property, to correspond.

In the case where all (other) arguments are \texttt{index} objects, when the \texttt{permute} method is invoked on a \texttt{tensor} object, a comparison is made between the \texttt{indices} property, the old indices, and the \texttt{index} arguments, the new indices. All old indices must be retained, variants unchanged. Additional new indices, which find trailing singleton dimensions, are allowed, in which case the \texttt{degree} increases. Each new \texttt{index} of a permuted \texttt{tensor} has the variant as specified.

Unlike \texttt{subsasgn}, \texttt{subsref} never invokes \texttt{permute}. A \texttt{subsref} operation with \texttt{index} subscripts ignores preexisting \texttt{indices} of the \texttt{tensor} operand. Concatenated \texttt{index} subscripts simply redefine \texttt{indices} of the returned \texttt{tensor}, notwithstanding contractions and attractions, which simplify \texttt{entries} and shorten \texttt{indices}, reducing \texttt{degree}.

The remaining unary methods of class \texttt{tensor}, operators included, may be divided into entrywise and pagewise groups. Entrywise unary methods, from \texttt{uminus} to \texttt{round}, apply a corresponding kernel function to the \texttt{entries} property. The \texttt{round} method, allowing a precision argument, illustrates a parameterized operation. Parameters are passed unchanged to the kernel function applied to the \texttt{entries} property.

Two pagewise methods, \texttt{transpose} and \texttt{ctranspose}, invoke kernel \texttt{pagetranspose} and \texttt{pagectranspose} functions on the \texttt{entries} of a \texttt{tensor}. These functions are like a \texttt{for} loop over third and additional dimensions of the \ac{MDA} where to each page, a contiguous 2D array of the first two dimensions, a kernel \texttt{transpose} or \texttt{ctranspose} is applied. Two other kernel functions for 2D arrays, \texttt{trace} and \texttt{diag}, were likewise extended for \acp{MDA} into \texttt{pagetrace} and \texttt{pagediag} functions. They are invoked on \texttt{entries} of a \texttt{tensor} by pagewise \texttt{trace} and \texttt{diag} methods.

\subsection{\emph{N}-ary Operations}
\label{sec:n-ary operations}

Kernel \texttt{horzcat} and \texttt{vertcat} functions compute column and row concatenations, respectively, for \acp{MDA}. These familiar $N$-ary operations, for vectors, matrices, and arrays of compatible dimensions, when overloaded by the \texttt{tensor} class as \texttt{horzcat} and \texttt{vertcat} methods, involve $N$ operands at once. Each overloaded method invokes a protected \texttt{alignn} method to modify each operand into a suitable \ac{MDA} prior to a \texttt{pagehorzcat} or \texttt{pagevertcat} invocation that produces the \texttt{entries} of the initial \texttt{tensor} result. The \texttt{alignn} method also supplies the \texttt{indices} of the initial result. The final \texttt{tensor} result is produced after contractions, specified by \texttt{alignn}, are computed using the \texttt{sum} method.

The \texttt{alignn} method inputs $N$ and outputs $N + 2$ arguments. For each \texttt{tensor} input only its \texttt{entries}, an \ac{MDA}, is output. Via a kernel \texttt{permute}, the \ac{MDA} is modified so that its dimensions align based on the set-theoretic \texttt{union}, in sequence, of \texttt{indices} irrespective of variants. A non-\texttt{tensor} input passes to an output only if it is a 2D array. One additional output is the set-theoretic \texttt{union} of \texttt{indices}, variants and left-to-right sequence preserved. If two variants of an \texttt{index} appear the leftmost is kept. The second additional output, an \texttt{index} vector, specifies which indices require summation, i.e., ones where two variants appeared.

To support outer concatenation, possible at non-row/column dimensions, \texttt{pagehorzcat} and \texttt{pagevertcat} invoke \texttt{pagecat} on the $N$ permuted \acp{MDA} from the \texttt{alignn} method. Corresponding non-singleton dimensions must equate in size except for the dimension at which concatenation is performed. As kernel \texttt{horzcat}, \texttt{vertcat}, and \texttt{cat} do not support outer concatenation, singleton dimensions in the third or higher position are expanded, using one kernel \texttt{repmat} per operand, to match expected sizes of corresponding dimensions after concatenation, now performable via a kernel \texttt{cat}.

Generalizing concatenation further, the \texttt{cat} method implements \texttt{index} concatenation. In its \texttt{InferiorClasses} list, the \texttt{tensor} class includes the \texttt{index} class so that the former's \texttt{cat} method is invoked in a \texttt{cat} expression where the first argument is an \texttt{index} object. The \texttt{cat} method converts the first argument, as required, into a number that specifies the dimension of concatenation, considering the \texttt{alignn} step. Apart from invoking \texttt{pagecat} directly, passing along the dimension of concatenation, the \texttt{cat} method is thereafter identical to the \texttt{horzcat} and \texttt{vertcat} methods.

Like \texttt{round}, the \texttt{isequal} method features in unit test details. Implementation of \texttt{isequal} is similar to \texttt{horzcat} and \texttt{vertcat}. However, the \texttt{isequal} method instead of a \texttt{tensor} returns a \texttt{logical}, equal to the result of a kernel \texttt{isequal}, also $N$-ary, invoked instead of \texttt{pagehorzcat} or \texttt{pagevertcat} on the \acp{MDA} from \texttt{alignn}. Moreover, instead of contractions the \texttt{isequal} method always returns \texttt{false} if one operand has an \texttt{index} of one variant and another operand has the same of the complementary variant. This result is congruous with how the kernel \texttt{isequal} of a column vector and its \texttt{transpose}, a row vector having identical entries in the same sequence, returns \texttt{false}.

In theory, MATLAB could invoke a single overloaded method for an $N$-ary product expression involving a \texttt{tensor}. However, it invokes a sequence of binary product methods with a left-to-right precedence rule. Prior to \ac{RT} software expression, this rule may require modification of an \ac{RT} algebraic expression. In particular, to express an $N$-ary inner product over an \texttt{index} all but the rightmost operand must have the same variant, \texttt{true} or \texttt{false}, for that \texttt{index}, with only the rightmost operand having a complementary variant.

The \ac{NT} software has two deviations from its \ac{NT} algebra. To correctly compute some expressions, one or more sub-expressions have to be enclosed in parentheses and preceded by an extra operator, whose purpose is to decrement internal counters associated with underline-equivalent operators on enclosed indices. For this purpose, the tilde operator is hijacked. Thus, it cannot represent the entrywise NOT of a \texttt{DenseNT}, a tensor class of the \ac{NT} software. These deviations represent additional weaknesses, this time from a constructivism angle, of \ac{NT} underline and multi-variant index formalisms.

\subsection{Binary Operations}
\label{sec:binary operations}

When the \texttt{subsasgn} method is invoked, for a \texttt{tensor} operand on the \ac{LHS} of an assignment, it provides parameters, namely subscripts of the \ac{LHS} operand, in addition to the \ac{RHS} operand. If all subscripts are \texttt{index} objects the assignment overwrites the \ac{LHS} operand with the \ac{RHS} one after its \texttt{permute} method is invoked so that \texttt{indices} are in the sequence specified by the \ac{LHS} subscripts and not the \ac{RHS} operand, required to be a \texttt{tensor}. If the \ac{RHS} operand is not a \texttt{tensor} here, a runtime error results. If all \ac{LHS} subscripts are numbers, the kernel's \texttt{subsasgn} function is applied to the \ac{LHS} \texttt{entries} with those subscripts and either the original \ac{RHS} operand, if not a \texttt{tensor}, or its \texttt{entries}.

Entrywise binary methods, from \texttt{plus} to \texttt{power} in Table~\ref{tab:tensorclass}, each invoke a protected method, \texttt{binary}, passing operands along as second and third arguments. The first argument is a handle, e.g., \texttt{@plus}, to an entrywise kernel function. Within \texttt{binary}, after it invokes a protected \texttt{alignn} on the operands, there are two \texttt{index} vectors and two \acp{MDA}. To compute initial \texttt{entries} of the final \texttt{tensor}, the entrywise kernel function is applied to the aligned \acp{MDA}. With the first \texttt{index} vector from \texttt{alignn} as initial \texttt{indices}, an initial \texttt{tensor} is produced using the two-argument constructor. Using the second \texttt{index} vector to specify contractions, the \texttt{sum} method is invoked to produce the final \texttt{tensor}.

Remaining binary methods, \texttt{mtimes}, \texttt{mldivide}, and \texttt{mrdivide}, are called pagewise because they exploit kernel \texttt{pagemtimes}, \texttt{pagemldivide}, and \texttt{pagemrdivide} functions through protected methods, namely \texttt{mbinary}, \texttt{mbinary2}, \texttt{align2}, and \texttt{lattice}. Each pagewise binary method invokes the \texttt{mbinary} method with a kernel function handle, e.g., \texttt{@pagemtimes}, as the first argument. For computational efficiency, the design involves a minimal number of kernel \texttt{reshape} and kernel \texttt{permute} invocations.

The pagewise binary methods exploit the \emph{lattice} concept. A lattice is a degree-one matrix represented as a 3D array. The matrix may be vector or scalar, depending on dimension sizes, and there may be one page. Lattices help to realize an arbitrary inner, outer, and entrywise \texttt{mtimes} between two \texttt{tensor} operands. Each operand maps (\texttt{permute}-and-\texttt{reshape}) to a lattice, $\mathbf{A}$ or $\textbf{B}$, a lattice product executes as follows, and the lattice result, $\mathbf{C}$, maps (\texttt{reshape}-and-\texttt{permute}) to a \texttt{tensor} with \texttt{indices} as required by the \ac{RT} algebra:
\begin{align}
\mathbf{C}_k &= \mathbf{A}_k \mathbf{B}_k \text{.}
\end{align}
All outer products are represented by rows of the first lattice, $\mathbf{A}$, and columns of the second lattice, $\mathbf{B}$. All inner products are represented by columns and rows, respectively, of the first and second lattices. The common page dimension represents all entrywise products, which may execute in parallel.

Similarly, the \ac{RT} software implements \texttt{mldivide} and \texttt{mrdivide} consistent with the \ac{RT} algebra. Such left- or right-divisions imply a solution to a corresponding linear system, allowing inner, outer, and entrywise products. These map to and from a left- or right-division of lattices, as follows:
\begin{align}
\mathbf{B}_k &= \mathbf{A}^k \backslash \mathbf{C}_k \text{,} \\
\mathbf{A}_k &= \mathbf{C}_k / \mathbf{B}^k \text{.}
\end{align}
As with scalar division, \texttt{indices} of denominator nonscalar operands have to change variants. Otherwise, the \ac{RT} framework would be inconsistent in the context of linear systems. Prior to invoking \texttt{mbinary}, the \texttt{mldivide} and \texttt{mrdivide} methods complement \texttt{indices} via a \texttt{not} operation.

The \texttt{mbinary} method receives the operands of a pagewise binary method as second and third arguments. If either operand is not a \texttt{tensor} then it must be a 2D array to avoid a runtime error by design. With a 2D array operand the \texttt{tensor} result has the same \texttt{indices} as the \texttt{tensor} operand. Resulting \texttt{entries} equal the result of a pagewise function, specified by handle, applied to the \texttt{tensor} operand's \texttt{entries} and the non-\texttt{tensor} operand. Because linear system operations are noncommutative, in general, pagewise function operands follow the original sequence received by \texttt{mbinary}.

When both operands of \texttt{mbinary} are \texttt{tensor} objects, the method invokes \texttt{mbinary2} on them. After \texttt{mbinary2} returns, \texttt{mbinary} gets from it the final \texttt{indices} of the operation, two \acp{MDA}, and two vectors that specify a kernel \texttt{reshape}-and-\texttt{permute}. To compute the final \texttt{entries} of the original operation, \texttt{mbinary} applies a pagewise function, specified via a handle, to the two \acp{MDA}, in the order received, followed by the kernel \texttt{reshape}-and-\texttt{permute}.

The \texttt{mbinary2} method invokes an \texttt{align2} method that depends only on \texttt{indices} of each operand. This results in matching and leftover \texttt{indices} for each operand. Additional vectors indicate whether matching \texttt{indices} have the same variants. After \texttt{align2} is done, \texttt{mbinary2} knows which dimensions of each operand's \texttt{entries} correspond in inner, outer, and entrywise fashion. Using this information, a kernel \texttt{permute}-and-\texttt{reshape} is applied separately, via the \texttt{lattice} method, to each operand's \texttt{entries}. Along with resulting \acp{MDA}, the \texttt{mbinary2} method returns information to \texttt{mbinary} for its kernel \texttt{reshape}-and-\texttt{permute}.

\begin{figure}[t]
\centering
\insertpdf{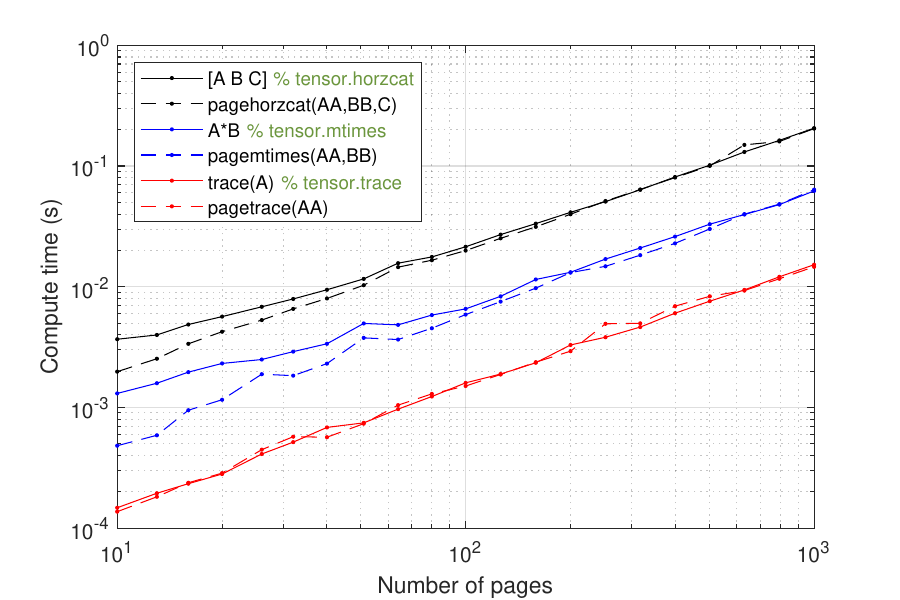}
\mycaption{Abstraction penalties vs.\ dimension size.}{The interpreted RTToolbox is computationally efficient when underlying unary, binary, or ternary operations involve large enough operands. In these tests, \texttt{tensor} operands have the same dimension sizes, $100 \times 100 \times P$. Tests vary just one dimension size, $P$.}
\label{fig:secsvspages}
\end{figure}

Figure~\ref{fig:secsvspages} presents representative abstraction penalties of \ac{RT} software, when operands are small enough, and computational efficiencies, when operands are large enough. In examples of unary, binary, and ternary operations with compatible dimension sizes, two operands, \texttt{A} and \texttt{B}, are degree-one matrices each constructed as a \texttt{tensor} with the same scalar \texttt{index}. Other operands, \texttt{C} or \texttt{AA} and \texttt{BB}, are 2D or 3D arrays.

For the Figure~\ref{fig:secsvspages} results, each operand or its \texttt{entries} is an array of class \texttt{double} with \texttt{rand} values. Compute times are averaged over multiple runs. Results are compared for equivalent operations done without \texttt{tensor} and \texttt{index} classes. In the binary case, the equivalent computation uses a kernel pagewise function. Although a kernel \texttt{mtimes}, for 2D arrays, may be applied to \acp{MDA} in pagewise fashion with one or more \texttt{for} loops (interpreted), the kernel \texttt{pagemtimes} (compiled) does the same but faster. As no kernel pagewise functions (interpreted or compiled) exist for the unary and $N$-ary cases, the comparisons use RTToolbox functions (interpreted).

The \ac{RT} software expresses the \ac{RT} algebra with programmatic efficiency. When abstraction penalties are small, relative to equivalent computations requiring only the kernel and novel pagewise functions, the \ac{RT} software achieves computational efficiency. Because kernel \texttt{ismember} and \texttt{unique} functions, critical to parsing the \ac{RT} algebra, are interpreted not compiled, they are a key source of penalties. So that the RTToolbox obeys toolbox recommendations of MATLAB Central File Exchange, it was designed without compilable MEX files, such as C/C++ implementations of kernel set-theoretic functions.

%%%%%%%%%%%%%%%%%%%%%%%%%%%%%%%%%%
% Conclusions
%%%%%%%%%%%%%%%%%%%%%%%%%%%%%%%%%%

\section{Conclusions}
\label{sec:conclusions}

This paper introduced an \ac{RT} framework, the successor to a published \ac{NT} framework, both of which complement a popular \ac{EMV} framework. Comprising an underlying \ac{RT} algebra and codesigned \ac{RT} software, to model a problem and develop a solution, the proposed \ac{RT} framework represents an approach for model-based imaging involving numeric tensors.

The \ac{RT} algebra inherits advantages of the \ac{NT} algebra, with respect to $N$-degree support, associativity, commutativity, entrywise products, linear invertibility, and other published considerations. However, due to its dual-variant index notation, as opposed to the \emph{de facto} multi-variant index notation of the \ac{NT} algebra, the \ac{RT} algebra provides all the expressivity of the \ac{NT} algebra in a simpler way. Thanks to additional outer operations, inspired by \ac{EMV} broadcasting, the \ac{RT} algebra is also more expressive than the \ac{NT} one. Although it aligns well with the Ricci notation of geometric tensor calculus, this paper proposes the \ac{RT} algebra for numeric tensor purposes.

Relevance of the \ac{RT} algebra for model-based approaches to imaging is illustrated with an example, which demonstrates an asymptotically-efficient solution of an optimization problem where the \ac{SSE}, a scalar function of deviant pixels in the image plane, depends on an unknown phase aberration in the pupil plane, a Fourier domain. The \ac{RT} algebra helps to model and simplify the \ac{SSE} and its first and second derivatives, namely a gradient and an \ac{HMF}, where the complexity of computing all three proves equivalent to computing just the \ac{SSE} alone. The approach makes visible diffraction rings and exoplanets near an occulted star in a simulated coronagraph image.

Thanks to the \ac{RT} algebra, all nonzero-degree matrices in a gradient formulation were eliminated. Moreover, gradient and \ac{HMF} expressions were reorganized to reveal (inverse) 2D \acp{DFT}, acceleratable via (inverse) 2D \acp{FFT}. Whereas the asymptotic time-and-space efficiencies were presented using MATLAB and not the \ac{RT} software, the latter proved useful. A natural implementation of the \ac{RT} algebra simplified a validation of intermediate steps in the underlying derivation. During the search for asymptotic efficiencies, algebraic errors were corrected via related \ac{RT} software sanity checks.

The \ac{RT} software is defined as the RTToolbox, comprising \texttt{tensor} and \texttt{index} classes, new pagewise functions, and unit tests, plus MATLAB. In this manner, the \ac{RT} framework complements a popular \ac{EMV} framework. Methods, functions, and tests presented, a subset of what could be developed, were chosen to emphasize key software requirements. Arguing that additional computational efficiency is possible, the approach taken emphasizes programmatic efficiency. Allowing the \ac{RT} algebra to be naturally expressed with MATLAB, the RTToolbox is designed in a way where small abstraction penalties, related to \texttt{index} objects, would reduce with a MATLAB release that compiled kernel set-theoretic functions.

Compared to the \ac{NT} software for MATLAB end users, comprising a C/C++ library called LibNT, M-file and MEX wrappers of the NTToolbox, plus MATLAB itself, the \ac{RT} software is simpler, comprising only M-files of the RTToolbox plus MATLAB. The toolbox uses an object-oriented approach to realize a dual-variant index notation, with its \texttt{index} class inheriting a kernel \texttt{handle} class, congruous to C/C++ \texttt{void} pointers, for extra programmatic and computational efficiency. In contrast, while the NTToolbox has a tensor class called \texttt{DenseNT}, it relies on MATLAB character vectors and string processing to realize a multi-variant index notation.

Finally, the paper elaborated on unary, binary, and $N$-ary \texttt{tensor} operations. For example, to express an $N$-ary inner product over an index, the \ac{RT} software may vary from the \ac{RT} algebra. As MATLAB parses an $N$-ary product into a binary product sequence with left-to-right precedence, all but the rightmost \texttt{index} must express an entrywise, not inner, product. This variation is contrasted with those of the \ac{NT} software, which unlike the \ac{NT} algebra required some products to be enclosed in parentheses and preceded with an additional operator. Other than introducing a left-to-right precedence rule, the \ac{RT} software expresses the \ac{RT} algebra faithfully.

%%%%%%%%%%%%%%%%%%%%%%%%%%%%%%%%%%
% Acknowledgements
%%%%%%%%%%%%%%%%%%%%%%%%%%%%%%%%%%

\section*{Acknowledgments}
\label{sec:Acknowledgments}

The author gratefully acknowledges editorial advice from Dan Sirbu, who suggested illustrating the \ac{RT} framework with a model-based example involving \acp{DFT}, and Adam P.\ Harrison, whose \ac{NT} framework laid a foundation for this work. He also thanks anonymous reviewers for constructive feedback that led to significant improvements, especially to clarity of the paper's topics and goals and to its imaging example.

%%%%%%%%%%%%%%%%%%%%%%%%%%%%%%%%%%
% Code and Data
%%%%%%%%%%%%%%%%%%%%%%%%%%%%%%%%%%

\section*{Code and Data}
\label{sec:code and data}

Code is shared with a permissive free software licence. For the imaging example, including its free source image, visit \href{https://www.github.com/KoderKong/Fourier2D}{www.github.com/KoderKong/Fourier2D}. For the RTToolbox, visit \href{https://www.mathworks.com/matlabcentral/fileexchange/156174-ricci-notation-tensor-toolbox-rttoolbox}{www.mathworks.com/matlabcentral/fileexchange/156174-ricci-notation-tensor-toolbox-rttoolbox}.

%%%%%%%%%%%%%%%%%%%%%%%%%%%%%%%%%%
% References
%%%%%%%%%%%%%%%%%%%%%%%%%%%%%%%%%%

\bibliography{RTFimaging}
\bibliographystyle{spiejour}

\end{document}